\documentclass[epj]{svjour}

\usepackage{graphicx}
\usepackage{epsfig}

\usepackage{amssymb}

\begin{document}

\title{Measurement of charged pions in $^{12}$C + $^{12}$C collisions at 1A~GeV and 2A~GeV with HADES}

\author{
G.~Agakishiev\inst{8}, 
C.~Agodi\inst{1},
A.~Balanda\inst{3,V},
G.~Bellia\inst{1,I},
D.~Belver\inst{15},
A.~Belyaev\inst{6},
J.~Bielcik\inst{4},
A.~Blanco\inst{2},
A.~Bortolotti\inst{9};
J.~L.~Boyard\inst{13}
% ...
\\
}

%
%

\institute{
\inst{1} Istituto Nazionale di Fisica Nucleare - Laboratori Nazionali del Sud, 95125~Catania, Italy \\
\inst{2} LIP-Laborat\'{o}rio de Instrumenta\c{c}\~{a}o e F\'{\i}sica Experimental de Part\'{\i}culas , 3004-516~Coimbra, Portugal\\
\inst{3} Smoluchowski Institute of Physics, Jagiellonian University of Cracow, 30-059~Krak\'{o}w, Poland\\
\inst{4} GSI Helmholtzzentrum f\"{u}r Schwerionenforschung, 64291~Darmstadt, Germany\\
\inst{5} Institut f\"{u}r Strahlenphysik, Forschungszentrum Dresden-Rossendorf, 01314~Dresden, Germany\\
\inst{6} Joint Institute of Nuclear Research, 141980~Dubna, Russia\\
\inst{7} Institut f\"{u}r Kernphysik, Johann Wolfgang Goethe-Universit\"{a}t, 60438 ~Frankfurt, Germany\\
\inst{8} II. Physikalisches Institut, Justus Liebig Universit\"{a}t Giessen, 35392~Giessen, Germany\\
\inst{9} Istituto Nazionale di Fisica Nucleare, Sezione di Milano, 20133~Milano, Italy\\
\inst{10} Institute for Nuclear Research, Russian Academy of Science, 117312~Moscow, Russia\\
\inst{11} Physik Department E12, Technische Universit\"{a}t M\"{u}nchen, 85748~M\"{u}nchen, Germany\\
\inst{12} Department of Physics, University of Cyprus, 1678~Nicosia, Cyprus\\
\inst{13} Institut de Physique Nucl\'{e}aire (UMR 8608), CNRS/IN2P3 - Universit\'{e} Paris Sud, F-91406~Orsay Cedex, France\\
\inst{14} Nuclear Physics Institute, Academy of Sciences of Czech Republic, 25068~Rez, Czech Republic\\
\inst{15} Departamento de F\'{\i}sica de Part\'{\i}culas, University of Santiago de Compostela, 15782~Santiago de Compostela, Spain\\
\inst{16} Instituto de F\'{\i}sica Corpuscular, Universidad de Valencia-CSIC, 46971~Valencia, Spain\\
\inst{I} Also at Dipartimento di Fisica e Astronomia, Universit\`{a} di Catania, 95125~Catania, Italy\\
\inst{II} Also at ISEC Coimbra, ~Coimbra, Portugal\\
\inst{III} Also at Universidade de Coimbra, ~Coimbra, Portugal\\
\inst{IV} Also at Dipartimento di Fisica, Universit\`{a} di Milano, 20133~Milano, Italy\\
\inst{V} Also at Panstwowa Wyzsza Szkola Zawodowa , 33-300~Nowy Sacz, Poland\\
}

\date{Received: \today / Revised version: \today}

\abstract{
We present the results of a study of charged pion production in $^{12}$C + $^{12}$C 
collisions at incident beam energies of 1A~GeV
and 2A~GeV using the HADES spectrometer at GSI. The main emphasis of
the HADES program is on the dielectron signal from 
the early
phase of the collision. Here, however,  we discuss the data with respect to the
emission of charged hadrons, specifically the production of
$\pi^\pm$ mesons, which are related to neutral pions
representing a dominant contribution to the dielectron yield.
We have performed the first large-angular range
measurement of the distribution of $\pi^\pm$ mesons for the 
$^{12}$C + $^{12}$C collision system
covering a fairly large rapidity interval.
The pion yields, transverse-mass and
angular distributions are compared with calculations done within
a transport model, as well as with existing data from
other experiments. The anisotropy of pion production 
is systematically analyzed.
 \PACS{
        {25.75.-q}{ heavy-ion collisions - }
        {25.75.Dw}{ charged pion spectra}
      } 
}

\titlerunning{Measurement of charged pions in $^{12}$C + $^{12}$C collisions at 
1A~GeV and 2A~GeV with HADES}
\authorrunning{G.~Agakishiev et al.}
\maketitle

\section{Introduction}

The investigation of nuclear matter at
high temperature and high density is one of the major research
topics in modern nuclear physics.  Nucleus-nucleus
collisions at relativistic energies offer the unique possibility to
create such highly excited nuclear matter in the laboratory 
\cite{ref1,ref1a,ref1b,ref1bb,ref1c,ref1d}.  The
study of particle production as function of beam energy, system size
and centrality of the collisions has been instrumental in the past
for understanding the approach of the system towards equilibrium and 
the generation of flow phenomena,
as well as for gaining information about the nuclear equation of state.
Collisions of the light $^{12}$C + $^{12}$C system represent a link
between the elementary nucleon-nucleon reactions and the heavy-ion collisions
of large nuclei.  Important physics issues in this context are the
degree of thermalization achieved, the role of the mean field
and collective motion in such a small system.

In the few-GeV energy range, pions are
the only abundantly produced mesons.  In heavy-ion collisions their spectra
and yields are affected by collective effects like thermalization,
directed and elliptic flow, as well as by possible modifications of the
properties of the baryon resonances they stem from, in particular
the $\Delta$ \cite{Mosel1,Mosel2}.
The subtle interplay of the phenomena which change the
characteristics of pion production with respect to nucleon-nucleon
interactions is indeed a challenge to theoretical interpretations.

Best suited for a description of all phases of the complex 
dynamics of heavy-ion reactions
are transport models, based on microscopic kinetic theory.
The reaction is simulated as a network of multiple elementary collisions
optionally embedded in a mean field equipped with
momen\-tum-dependent potentials.
These assumptions are then tested by comparing the experimental observables 
with the model predictions and allow to an understanding of the reaction 
dynamics. 
Transport models achieve a remarkable success in describing bulk properties
of the interactions over a large energy and system size scale (cf. 
\cite{ref1,ref1a,ref1b,ref1bb,ref1c,ref1d}).
At the same time, however, for special channels, problems are met in reproducing
precisely the experimental data.  For a recent comprehensive discussion 
of various
differential pion observables and their comparison with model calculations
in the region of 1A~GeV see \cite{reisdorf}.

The~~High~~Acceptance~~ DiElectron~~Spectrometer (HADES) \cite{ref2}, in
operation at the heavy-ion synchrotron SIS18 at GSI, Darmstadt, is
designed for high-resolution and high-acceptance dielectron
spectroscopy in hadron-hadron, ha\-dron-nucleus, and nucleus-nucleus
reactions at beam energies in the range from 1A~GeV to 2A~GeV.  Being a
charged-particle detector, it is of course also an efficient device for
hadron detection.  First results from HADES on dielectron production in
$^{12}$C~+~$^{12}$C reactions have been presented in \cite{ref3,ref3a}.
The interpretation of the dilepton spectra relies on a precise knowledge of the differential yields of neutral
pions which are the source of the bulk of the detected dielectron pairs,
namely via the $\pi^{0}$ Dalitz and photon decays.  In these analyses,
the $\pi^{0}$ yields were inferred from the charged-pion yields measured by
HADES in the same $^{12}$C~+~$^{12}$C data samples.

In this paper we present detailed data on charged pions
obtained from $^{12}$C + $^{12}$C collisions at 1A~GeV and 2A~GeV.  For the
first time, large intervals of rapidity ($\approx \pm0.8$ in $y/y_{beam}$ 
for 2A~GeV) and of
centre-of-mass angle ($-0.7 < \cos(\theta_{cms}) < 0.7$) are covered.
Our results are compared to UrQMD transport-model predictions and 
experimental data from other experiments.

\section{Experiment}

\begin{figure}[htb]
  \vspace*{+.5cm}
\includegraphics[width=1.\columnwidth]{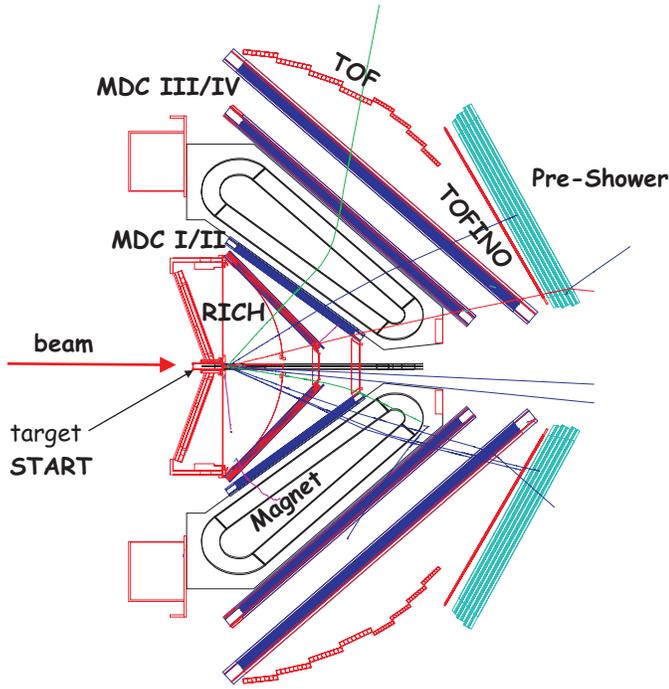}
  \vspace*{-0.1cm}
  \caption[]{Cut through two sectors of the HADES spectrometer.
  The magnet coils are projected onto the cut plane to visualize 
  the toroidal magnetic field. 
  The average distance
  between the target and the outermost detectors is about 210 cm.}
  \label{had1}
\end{figure}

HADES \cite{ref2} is a magnetic spectrometer designed as
second-generation device for measurements of $e^+e^-$ pairs. The
spectrometer, schematically depicted in Fig.~\ref{had1}, is
segmen\-ted into six identical sectors that cover laboratory polar angles between
18 and 85 degrees. Its large azimuthal acceptance
covers between 65\% and 90\% of 2$\pi$ at small and large polar
angles, respectively. The analysis of charged pions presented here
is based on the same setup as used in the lepton analyses of 
\cite{ref3,ref3a}. A fast
hadron-blind Ring Imaging CHerenkov counter (RICH) is used for electron and
positron identification. In the pion analysis, the RICH information was 
not used.
Four planes of Multi-wire Drift Chambers
(MDC I - MDC IV), together with a superconducting magnet, form the
magnetic spectrometer for track reconstruction and momentum
determination. In the region behind the magnetic field, a set of
electromagnetic Pre-Shower detectors (at polar angles $18^\circ - 45^\circ$)
\cite{presh} and a time-of-flight wall \cite{tof} are installed
which form the META (Multiplicity and Electron Trigger Array). The
time-of-flight detector wall is subdivided into 2 regions: TOF (at
polar angles $45^\circ - 85^\circ$), consisting of 384 scintillator slabs of
varying length, which are read out at both ends, with a
time-of-flight resolution of $\sigma$ = 150 ps, and TOFINO (at polar
angles $18^\circ - 45^\circ$), consisting of 24 scintillator plates
readout on one end, with a time-of-flight resolution of $\sigma$ =
450 ps. 
The TOFINO is placed directly in front of the Pre-Shower detector, which
provides precise position measurement.
The TOF/TOFINO detectors are also used for fast charged-particle
multiplicity measurements. Together with the Pre-Shower detectors
they provide additional lepton/hadron discrimination power and track
coordinate measurements with a spatial resolution in the range from
14 to 25~mm.

A fast data acquisition system is used together with a two-level trigger
scheme \cite{ref9a,ref9b}: (a) LVL1 is based on a fast determination of the
charged-particle multiplicity ($M_{ch}$) in the time-of-flight detectors.
(b) LVL2 is based on a real time identification of electron and
positron candidates. All LVL2-triggered events were written to tape,
as well as a part of LVL1 (regardless of the LVL2 decision) events 
(typically 10\%) for normalization
purposes, hadron analysis and monitoring of the trigger performance.
For the analysis presented here, only LVL1-triggered events
were processed.

In the very first HADES physics run, the detector was operated using only
the following sub-systems: the RICH, the two inner MDC planes, and the META,
i.e. the two outer MDC planes were not operational and the position
measurements from the META were used for tracking.
In this mode, the collision system $^{12}$C~+~$^{12}$C at 2A~GeV was
studied with a beam intensity of $I_{beam} = 10^6$ particles/sec
impinging on a two-fold segmented carbon target with thickness of 
$2 \cdot 2.5\%$
interaction length.
$1.67 \cdot 10^7$ LVL1 triggered events with $M_{ch} \ge 4$ were analyzed in this
study.
In the event reconstruction, the track segments measured in the two inner MDC
planes were correlated with hits in the META.

In the second data taking period the $^{12}$C~+~$^{12}$C system was studied at 1A~GeV.
Then, for the first time, a high-resolution tracking mode exploiting
also the outer MDC planes was available. In this measurement, a carbon beam of
$10^6$ particles/sec was focused onto a carbon foil of 3.8\% interaction length.
$1.62 \cdot 10^7$ LVL1-triggered events with $M_{ch} \ge 4$ were used in this
analysis.

\section{Data Analysis}

\subsection{Simulations}

Artificial $^{12}$C + $^{12}$C events were generated with the UrQMD (v1.3b) 
transport code \cite{ref10,ref11}. The detector response was simulated with
the help of a GEANT 3.21 based package \cite{ref12} including the geometry and
characteristics of all HADES detectors. Then the same LVL1 trigger condition 
($M_{ch}~\ge~4$) was applied. The resulting raw data were processed
in exactly the same way as the real data and used for efficiency corrections 
as well as to estimate systematic errors. Details on the different
procedures are given in the corresponding subsections.
For 1A GeV we have analyzed $2.14 \cdot 10^7$ (LVL1) UrQMD events, for 2A GeV 
$2.07 \cdot 10^7$ events, i.e. numbers being comparable to the amount
of analyzed real data. 
For this sample size, the statistical errors of the $\pi$ yields in the 
region of 
interest are negligible.
Systematic errors  are not addressed by the UrQMD itself, but may be
elucidated by comparison with data. For the description of the apparently
complex process of pion production 
in heavy-ion reactions in this energy range, the precision of transport
models is not expected
to be better than 20\% \cite{ref1bb}.

Additionally, we simulated the $\pi$ meson emission using a simple 
Monte-Carlo event generator PLUTO \cite{pluto}, which
assumes a thermal source modified by a polar angular distribution. 
The simulation parameters used for PLUTO - inverse slopes and
anisotropies - were derived from our measured data. The generated 
rapidity distribution has been used to extrapolate the pion yields
outside our detector acceptance. Varying the input parameters of the generator
within their errors provides estimates for the  
systematic error of the extrapolation.

\subsection{Momentum reconstruction}

When traversing the spectrometer, charged particles are deflected in
the magnetic field and at the same time they leave "hits" in the
MDCs and META detectors. From this information, together with the
known magnetic field, their trajectories are reconstructed and their
momenta are deduced.

Two different tracking methods have been developed and were both
used in the present analysis (see \cite{ref2} for details). The
first one is the ``kickplane'' algorithm which uses the position
information delivered by the inner MDC chambers and the META system.
In this case the momentum resolution $\sigma_{p}/p$, dominated by
the limited position resolution of META, has been determined in
simulations to be $\simeq 2.5\%$ at a momentum of 150 MeV/c, with a
linear increase up to $16\%$ at 1400 MeV/c, with a weak dependence 
on polar angle. 
The second method is based on a
Runge-Kutta trajectory integration routine \cite{RK} which uses
the information from all four MDC planes, resulting in a resolution
$\sigma_{p}/p \simeq$ 1.8\% at 150 MeV/c and $\simeq$ 4.3\% at 1400 MeV/c. 
To obtain the results presented in this paper,
the kickplane method has been applied to the 2A~GeV data, and both
methods were used and compared for the 1A~GeV data.

\subsection{Particle identification}

\begin{figure*}[!ht]
  \vspace*{+.5cm}
  \includegraphics[width=1.8\columnwidth]{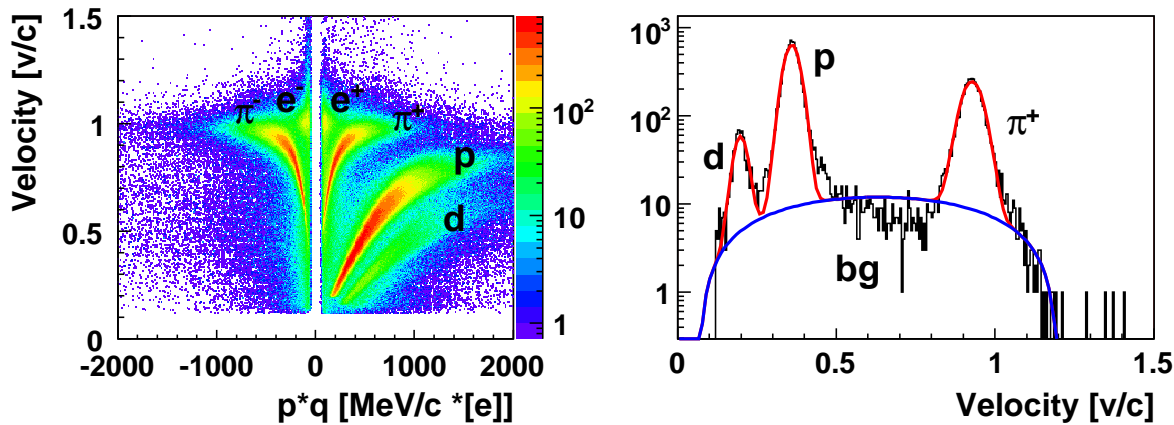}
  \vspace*{-0.1cm}
  \caption[]{Velocity vs.\ charge-times-momentum of charged particles as seen by the
   HADES detector from $^{12}$C + $^{12}$C collisions at 2A~GeV (left). Projection onto the
   velocity axis of positively charged particles with momenta 350$\pm$5~MeV/c and
   $\theta$~=~60$^\circ\pm$5$^\circ$ (right). Fitted signal and background distributions
   are shown as lines.}
  \label{pid1}
\end{figure*}

Particle identification in the HADES data analysis (for details see
\cite{ref2}) is based on Bayesian statistics \cite{Bayes,barlow}. The method
allows to evaluate the probability that the reconstructed track can
be related to a certain particle species (e.g. proton, kaon, pion, 
electron, etc.). It combines several observables from various
sub-detectors (e.g. time-of-flight, energy loss) via probability
density functions (PDF) determined for each observable and for all
possible particle species. The probabilities for different mass
assignments of any given track are calculated from the assumed
abundances of the individual particle species and from the PDFs of
all measured variables. The latter ones are obtained from
simulations. If the assumed abundances differ significantly from the
final results, the procedure is repeated with updated input
distributions. It converges typically after one or two iterations. The
performance of the method in terms of efficiency and purity is
evaluated in detailed simulations and simultaneous comparisons with
the real data. 
In our case, hadron identification has been performed
using measured momenta and corresponding velocities computed by means of the
time-of-flight. 
For more sophisticated analyses, like electron or rare-hadron identification,
data from the RICH and PreShower detectors as well as the energy
loss in META and MDCs can be used in addition.


The method used for Particle IDentification (PID) is illustrated in
Fig.~\ref{pid1} for the case of particle velocity (right) deduced from the
measured time-of-flight and track length (``velocity-vs-momentum''
algorithm). Particles with different mass occupy different regions
in the velocity-vs-momentum distribution (left side); the pronounced
ridges correspond to positively and negatively charged pions, protons and
deuterons. 
The very rare kaons are not visible in this representation.
The Bayesian PID method requires the determination of the
probability density functions for each particle species. In the
case of the velocity-vs-momentum algorithm used here, the PDF is the
probability distribution of velocity. For each type of particle it
has been determined in bins of momentum and polar angle. In those
velocity distributions Gaussian fits were used to obtain the signal
(i.e. particle yields) and a 2$^{nd}$-order polynomial fit to obtain
the background (due to fake tracks). The fitted distributions were
normalized to unity. Fig.~\ref{pid1} (right side) shows as an
example such a fit for the momentum bin 350$\pm$5~MeV/c in the polar angle range
$\theta$ = 60$^\circ\pm$5$^\circ$.

Two quality parameters are used to characterize the performance of
the method \cite{hommez}: the PID efficiency and the PID purity. The
PID efficiency $\varepsilon_{t}(p,\theta)$ is the probability that a
particle with the true type $t$ is identified as type $t$. The PID
purity $\pi_{t}(p,\theta)$ is the probability that a particle that
is identified as type $t$ is truly of type $t$. The PID efficiency
and purity have been studied in detailed simulations with events
generated with the UrQMD model. The critical parameter here is the
time resolution, which is however well known. Because of the moderate time
resolution of the presently installed TOFINO detectors this limits the
region in which we can use the method for $\pi^{+}$ and p
identification to momenta $<1000$ MeV/c . Finally, we have 
checked that varying
particle abundances by a factor as large as 2 does not change the results
significantly.

\begin{figure*}[!ht]
  \vspace*{+.5cm}
  \includegraphics[width=1.8\columnwidth]{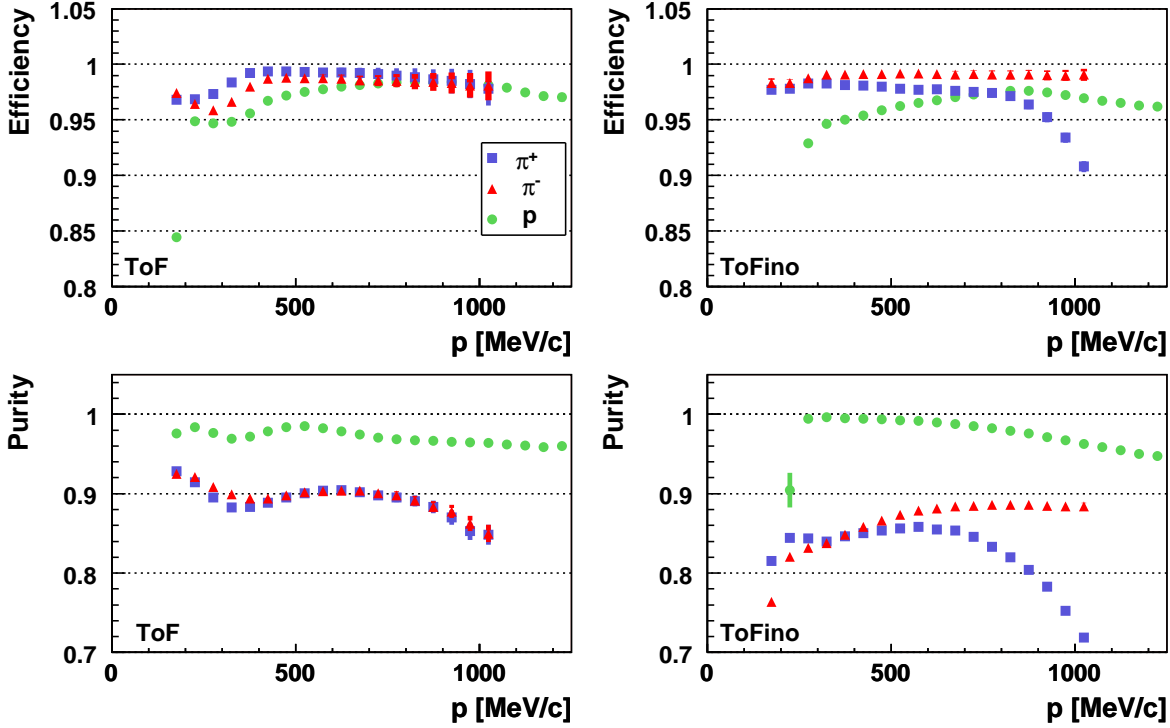}
  \vspace*{-0.1cm}
  \caption[]{Efficiency (top) and purity (bottom) of the PID method
   versus momentum for $\pi^\pm$ and protons in two sub-systems of the
   HADES detector: TOF (left) and TOFINO+PreShower (right) by using the kickplane
   reconstruction algorithm in $^{12}$C~+~$^{12}$C collisions at 2A~GeV.}
  \label{pid2}
\end{figure*}

\subsection{Total correction}

Fig.~\ref{pid2} shows the dependencies of the PID efficiency and
purity on momentum for $\pi^+$, $\pi^-$ and protons for the 2A~GeV
data in the TOFINO (right) and TOF (left) region. The efficiency of
pion and proton identification is larger than 95\% for all momenta
in the TOF region. In the TOFINO region, with its reduced time
resolution, the efficiency to identify positively charged pions drops
steeply above 1000 MeV/c due to the ambiguity with the protons. The
purity of pions (lower plots) does not reach unity mainly because
part of the tracks identified as pions are muons from
in-flight pion decays. The strong contamination of the positively
charged pions with protons for momenta above 1000 MeV/c is again
due to the low time resolution of the TOFINO.

After the particle identification is done for all tracks, the
resulting yields are corrected for efficiency and purity of the PID
method, as well as for the detector and tracking efficiencies. The
detection/tracking efficiency has also been obtained from Monte
Carlo simulated and reconstructed UrQMD events. 
The reliability of the simulation was cross checked with real data by
comparing and matching the information from different detectors and by
selecting clean samples of (elastic) elementary collisions in which the
track configurations can be obtained from the trigger detector (META) 
information only \cite{ref2}. Using this information, realistic parameters were 
validated for the simulation.

The total correction
applied to the reconstructed particle yields reads
\begin{equation}
w_t(p,\theta) = \frac{\pi_{t}(p,\theta)}
{\varepsilon_{t}(p,\theta)~\varepsilon_{t}^{det}(p,\theta)} \label{eq:effcor},
\end{equation}
where $\varepsilon_{t}^{det}(p,\theta)$, the detection efficiency,
subsumes detector, track reconstruction and acceptance losses. It
should be noted that we specify $\varepsilon_{t}^{det}(p,\theta)$ as
function of $\theta$ and $p$, while averaging over the azimuthal
angle. In this way, corrections for missing geometrical acceptance
at some azimuthal angles (namely the spaces occupied by the six
magnet coils) are accounted for as well as the losses due to pion
in-flight decays.

\begin{figure}[htb]
  \vspace*{+.5cm}
  \includegraphics[width=1.\columnwidth]{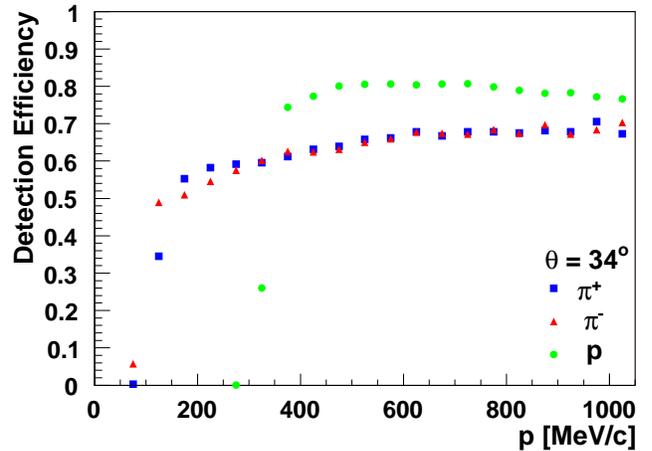}
  \vspace*{-0.1cm}
  \caption[]{The detection efficiency $\varepsilon_{t}^{det}(p,\theta)$ vs.\ momentum
  for $\pi^\pm$ and
   protons using the kickplane reconstruction in $^{12}$C+$^{12}$C collisions at 2A~GeV.

}
  \label{accVsMom}
\end{figure}

Figure~\ref{accVsMom} shows the dependence of the detection efficiency
for $\pi^\pm$ and p as a function of momentum. The difference
between proton and $\pi$ efficiencies is again in part caused by the
$\pi^\pm$ in-flight decay.

The total correction is applied to the data for each momentum and
polar angle bin, and for each individual particle species. This is
done only for bins with sufficiently high efficiency
$\varepsilon_{t}^{det}(p,\theta)>0.35$ in order to avoid large
corrections at the sector boundaries. Data outside of this fiducial
volume were excluded from further analysis.

 \begin{figure}[ht]
  \includegraphics[width=1.\columnwidth]{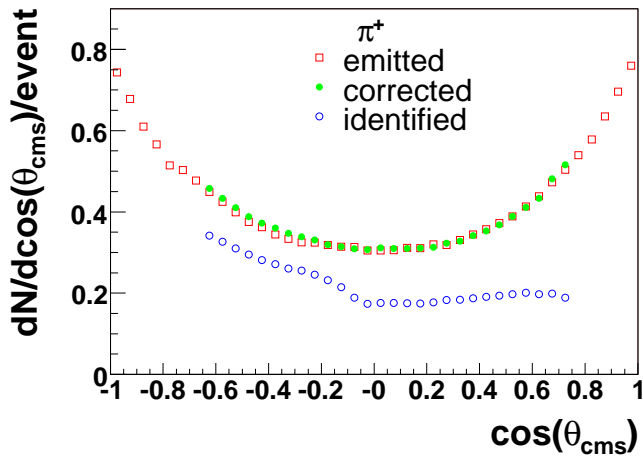}
  \vspace*{-0.1cm}
  \caption[]{Center-of-mass polar angle distributions of $\pi^+$.
The squares represent the pions as generated by UrQMD. Only $\pi$
mesons with $p_{cms}>200~{\rm MeV/c}$ have been selected. The open circles
show those generated pions which are detected and identified in the
HADES acceptance. The full circles depict the result of the efficiency
and purity correction to the accepted simulated pions. }
  \label{pid3cm}
\end{figure}

Figure~\ref{pid3cm} presents simulated polar distributions of pions
in the center-of-mass system (cms). It shows the identified
$\pi^+$ before and after applying the total correction, together with
the primordial distribution delivered by the UrQMD model for 2A~GeV 
$^{12}$C~+~$^{12}$C collisions.
This self-consistency check quantifies $w_t(p,\theta)$ as a function of 
$\cos \theta_{cms}$
and demonstrates the wide coverage of our spectrometer. In Fig.~\ref{pid3cm} 
the angular anisotropy of pion emission in the UrQMD
model is clearly visible.

The point-to-point systematic errors of the measured distributions 
stem mainly from imperfect modelling of the 
detectors, reconstruction and identification efficiency. They are estimated 
as $\approx 5\%$, based on a comparison of measurements 
in the six independent HADES sectors. 
It should be noted that although the
six sectors are identical in design, they are different at small scales
like malfunction of electronic channels and small mechanical displacements.
The resulting errors are of statistical nature and thus appropriate for
fits to the various distributions.
The geometry of the chambers and the magnetic field measurement have some 
influence on high resolution tracking, but do not affect continuum 
distributions as those discussed in this work.

\subsection{Event selection}

\begin{figure*}[!htb]
  \vspace*{+.5cm}
  \includegraphics[width=1.8\columnwidth]{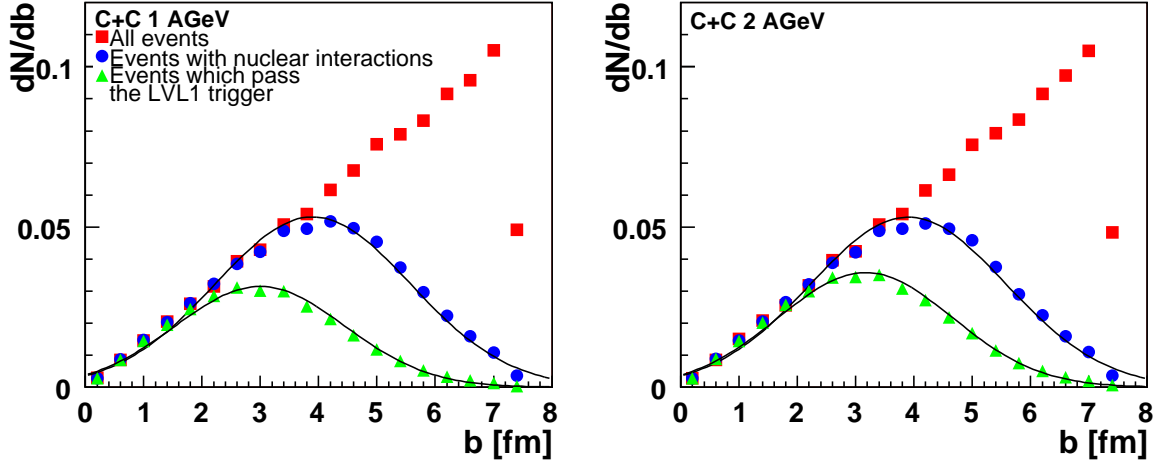}
  \vspace*{-0.1cm}
  \caption[]{The impact parameter distribution obtained from the UrQMD model  
for $^{12}$C~+~$^{12}$C collisions at 1~(left) and 2A~GeV (right).
}
  \label{imp}
\end{figure*}

For the present analysis, we used the HADES LVL1-trig\-gered events,
which are characterized by a hit multiplicity M$_{ch}~\ge~4$ in the
time-of-flight detectors. The correlation between the LVL1 trigger
condition and the centrality of the reaction has been studied in
Monte-Carlo simulations using the UrQMD and GEANT codes.
Fig.~\ref{imp} shows the simulated impact-parameter distributions. For
the ``minimum bias'' events corresponding to the total reaction
cross section we require in UrQMD at least one nuclear interaction
(distribution marked by circles in Fig.~\ref{imp}). Then we pass
these events through our analysis code and require that they fulfill
the LVL1 condition (triangles in Fig.~\ref{imp}). 
We found that the
LVL1-triggered events correspond to 52\% and 60\% of the total
reaction cross section in $^{12}$C + $^{12}$C collisions at 1A~GeV and
2A~GeV, respectively. To extract the average number of participants 
for our trigger-biased
events we proceeded in the following way: For minimum-bias events
the average number of participating nucleons was estimated from a
geometrical model \cite {gosset}, namely
$\langle $A$_{part}^{m.b.} \rangle $ = $A/2$ = 6. 
For reactions accepted by LVL1 we deduce the mean 
$\langle $A$_{part} \rangle$  by comparing the pion multiplicity of UrQMD 
for LVL1-accepted events 
to the one of minimum-bias events, using  
$\langle $A$_{part} \rangle$ scaling of 
pion production (see e.g. figure 32 in \cite{ref1a}), i.e. 
$\langle $A$_{part} \rangle = 6 \cdot \langle M_{\pi}^{LVL1} \rangle / \langle M_{\pi}^{min.b.} \rangle$.
The LVL1 trigger effect is significant, and the number of participants
increases by $\approx 40\%$. The average impact parameters, the
average pion multiplicities and average number of participants from
UrQMD are listed in Table~\ref{tab4} for true minimum-bias
events and after applying the LVL1 trigger.

\begin{table}[htb]
\begin{center}
\caption{Average impact parameters, pion multiplicities, 
and average number of participating nucleons
from UrQMD calculations for $^{12}$C~+~$^{12}$C at 1A~GeV and~2A~GeV 
before and after applying the LVL1
trigger condition. \label{tab4}}

\begin{tabular}{l c c c c c}
\\
\hline
\multicolumn{5}{c}{Beam energy = 1A~GeV} \\
\hline
 & $\langle b \rangle$ & $\langle M_{\pi^{+}} \rangle$ & $\langle M_{\pi^{-}}\rangle$ & $\langle $A$_{part}\rangle$ \\
minimum-bias events & 3.95 fm& 0.36 & 0.36 & 6 \\
LVL1-triggered & 3.01 fm& 0.51 & 0.52 & 8.61 \\
\hline
\multicolumn{5}{c}{Beam energy = 2A~GeV} \\
\hline
 & $\langle b \rangle $ & $\langle M_{\pi^{+}} \rangle $ & $\langle M_{\pi^{-}}\rangle $ & $\langle $A$_{part}\rangle$ \\
 minimum-bias events & 3.95 fm & 0.83 & 0.83 & 6 \\
LVL1-triggered & 3.18 fm & 1.15 & 1.17 & 8.38 \\
\hline
\end{tabular}\\
\end{center}
\end{table}

\begin{figure*}[htb]
   \vspace*{+.2cm}
  \includegraphics[width=1.8\columnwidth]{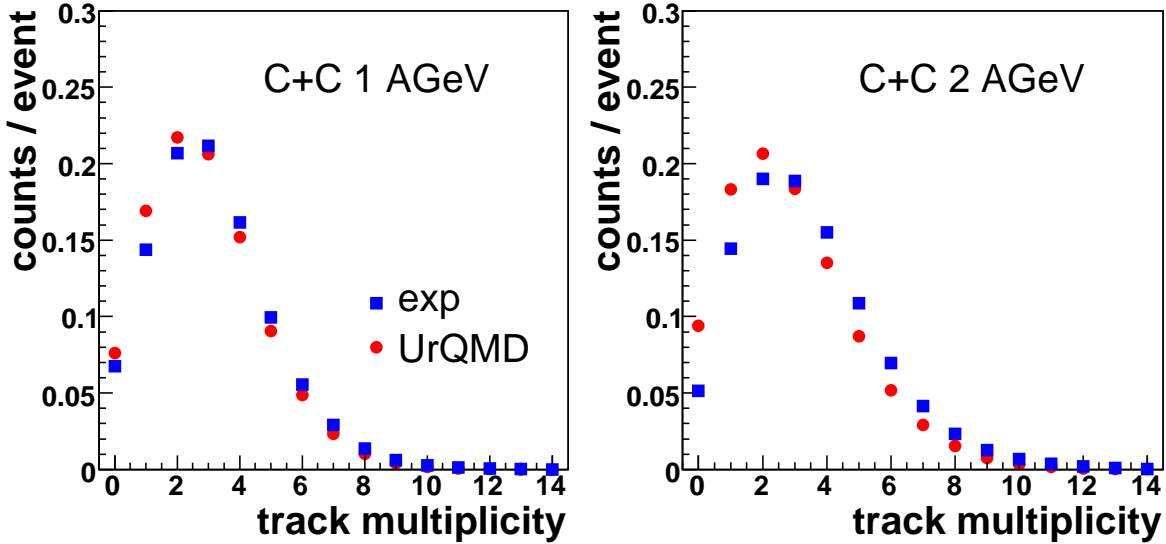}
   \vspace*{-0.1cm}
\caption[]{Distribution of the number of reconstructed tracks
 in the data and in the simulation for 1A~(left) and 2A~GeV (right) 
$^{12}$C~+~$^{12}$C collisions.}
\label{mult}
\end{figure*}

The distributions of the number of reconstructed tracks per LVL1 event of
data and UrQMD simulations are in reasonable agreement, as shown in 
Fig.~\ref{mult}.
This is a confirmation that the modelling of the detector and tracking
efficiency as well as of the LVL1 event selection in our simulation is
realistic. From the differences of the measured and simulated distributions 
of the number of charged hits in META and the number of reconstructed tracks 
we estimate a systematic error for the mean number of participants 
determined above, resulting in 
$\langle $A$_{part} \rangle = 8.61\pm{0.60}$ and $8.38^{+1.17}_{-0}$ for 1A~GeV and 2A~GeV, 
respectively. 

Comparing the pion momenta and angular distributions for minimum-bias 
and LVL1 UrQMD events is an important check for how the trigger condition 
used in the experiment influences the measured spectra. We do not observe 
significant deviations between the two data sets, e.g. for the transverse mass
distribution, the difference in the inverse slopes (see Section 4.1)
is less than 1 MeV, and the $A_2$ parameters of the polar angle distributions
(see Section 4.4)
differ by less than 10\%, while their momentum dependencies are identical. 
The reason is that for a small system such as $^{12}$C + $^{12}$C it is
basically impossible to select events with well defined centrality by a 
charged-particle 
multiplicity trigger. This fact is clearly seen in Fig.~\ref{imp}, which shows
a very broad distribution in impact parameter from the LVL1 events.  

\section{Results}

\subsection{Transverse-mass distributions}

 \begin{figure*}[htb]
  \vspace*{+.01cm}
  \includegraphics[width=1.8\columnwidth]{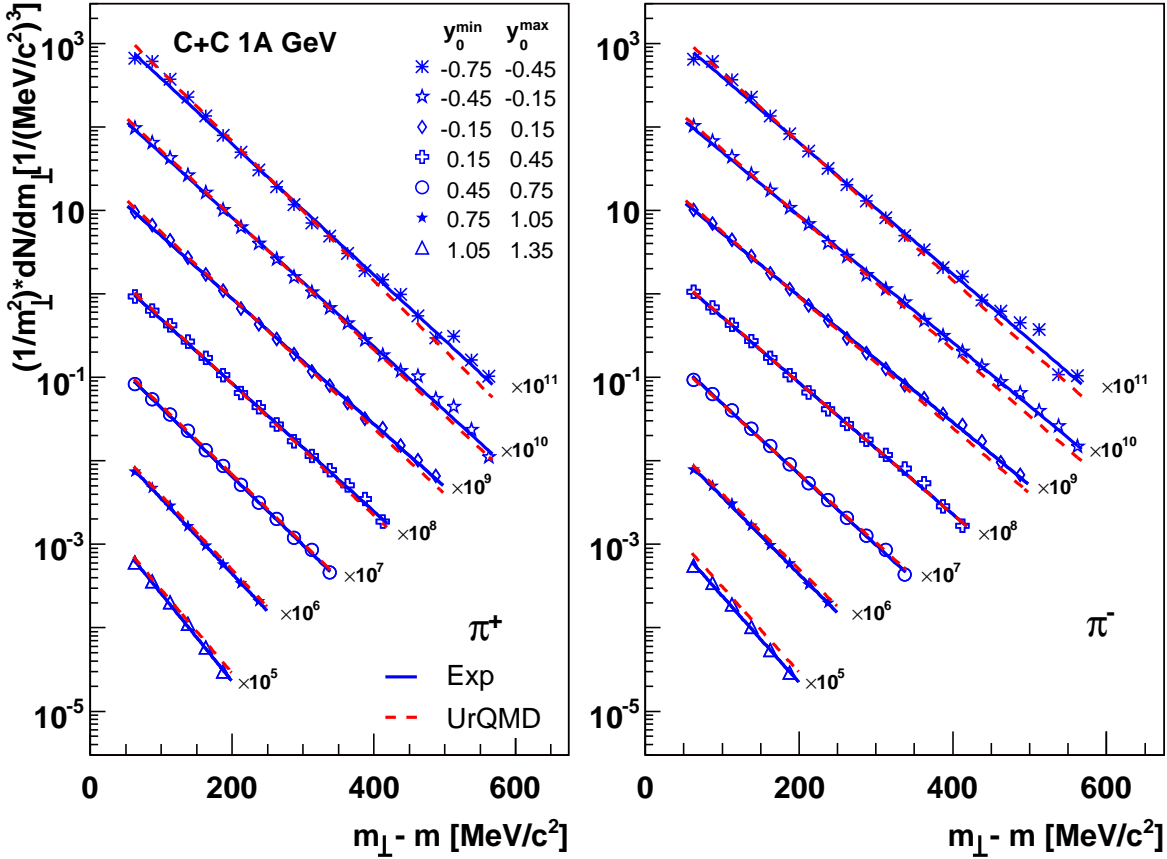}
  \vspace*{-0.1cm}
  \caption[]{Transverse-mass distributions for positively (left) and negatively
  (right) charged pions in different slices of rapidity derived from the data
  in $^{12}$C~+~$^{12}$C collisions at 1A~GeV (LVL1 ``semicentral'' events).
  Full lines show the results of fits to the data (symbols) using one 
  exponential
  function, while dashed lines show fits of the UrQMD distributions using 
  the same fit function. Error bars (systematic and statistical ones) are not
  visible at this scale. Both data and UrQMD distributions are normalized
  to the number of LVL1 events.}
  \label{dNdMpion_aug04rk}
\end{figure*}

 \begin{figure*}[htb]
  \vspace*{+.01cm}
  \includegraphics[width=1.8\columnwidth]{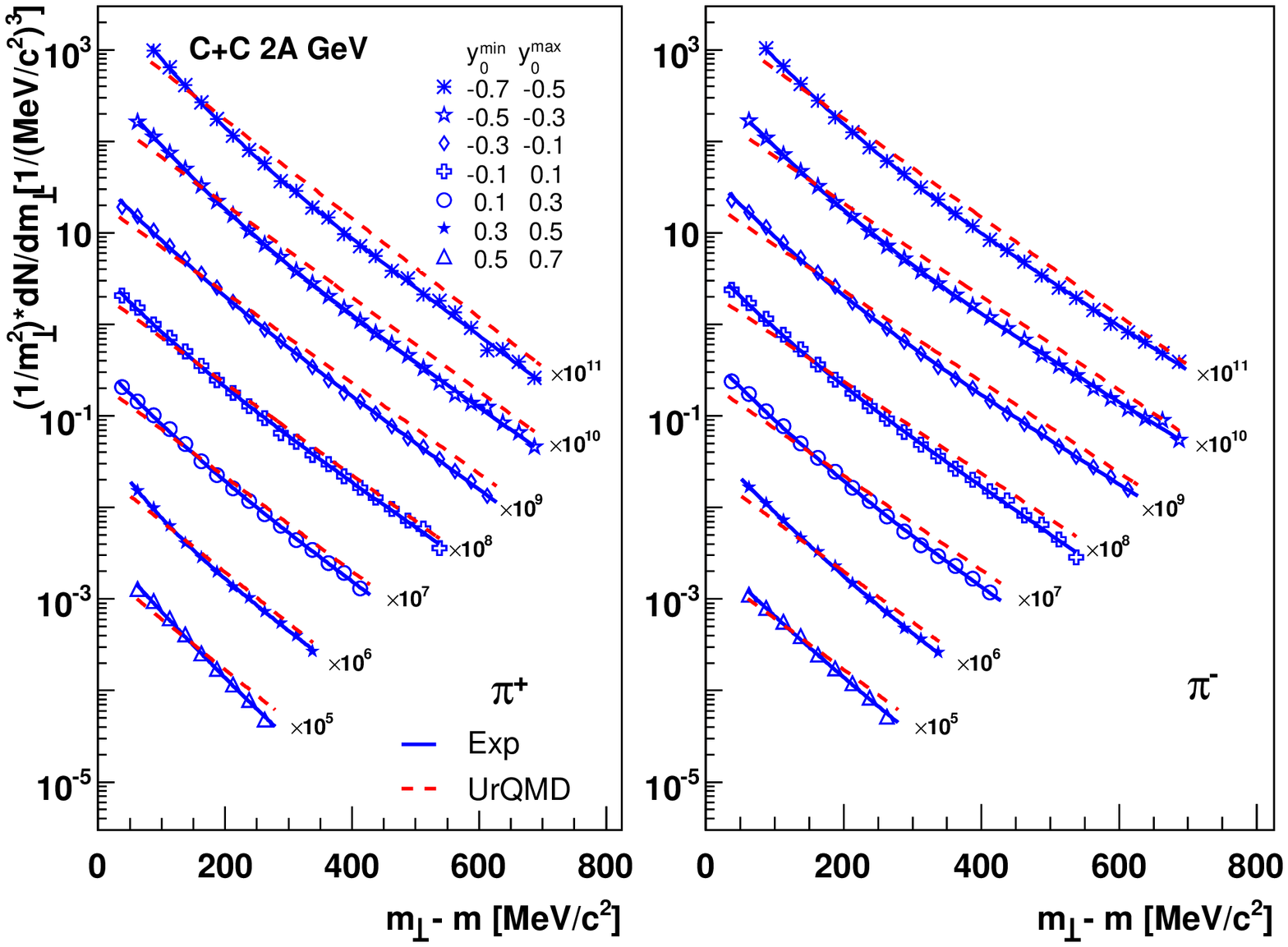}
  \vspace*{-0.1cm}
  \caption[]{Transverse-mass distributions for positively (left) and negatively
  (right) charged pions in different slices of rapidity derived from the data
  in $^{12}$C~+~$^{12}$C collisions at 2A~GeV (LVL1 ``semicentral'' events). 
  Full lines show the results of fits to the data (symbols) using two 
  exponential 
  functions, while dashed lines show fits of the UrQMD distributions using 
  one exponential function. Error bars (systematic and statistical ones) are not
  visible at this scale. Both data and UrQMD distributions are normalized
  to the number of LVL1 events.}
  \label{dNdMpion}
\end{figure*}

\begin{center}
\begin{figure*}[!ht] 
  \includegraphics[width=1.\columnwidth]{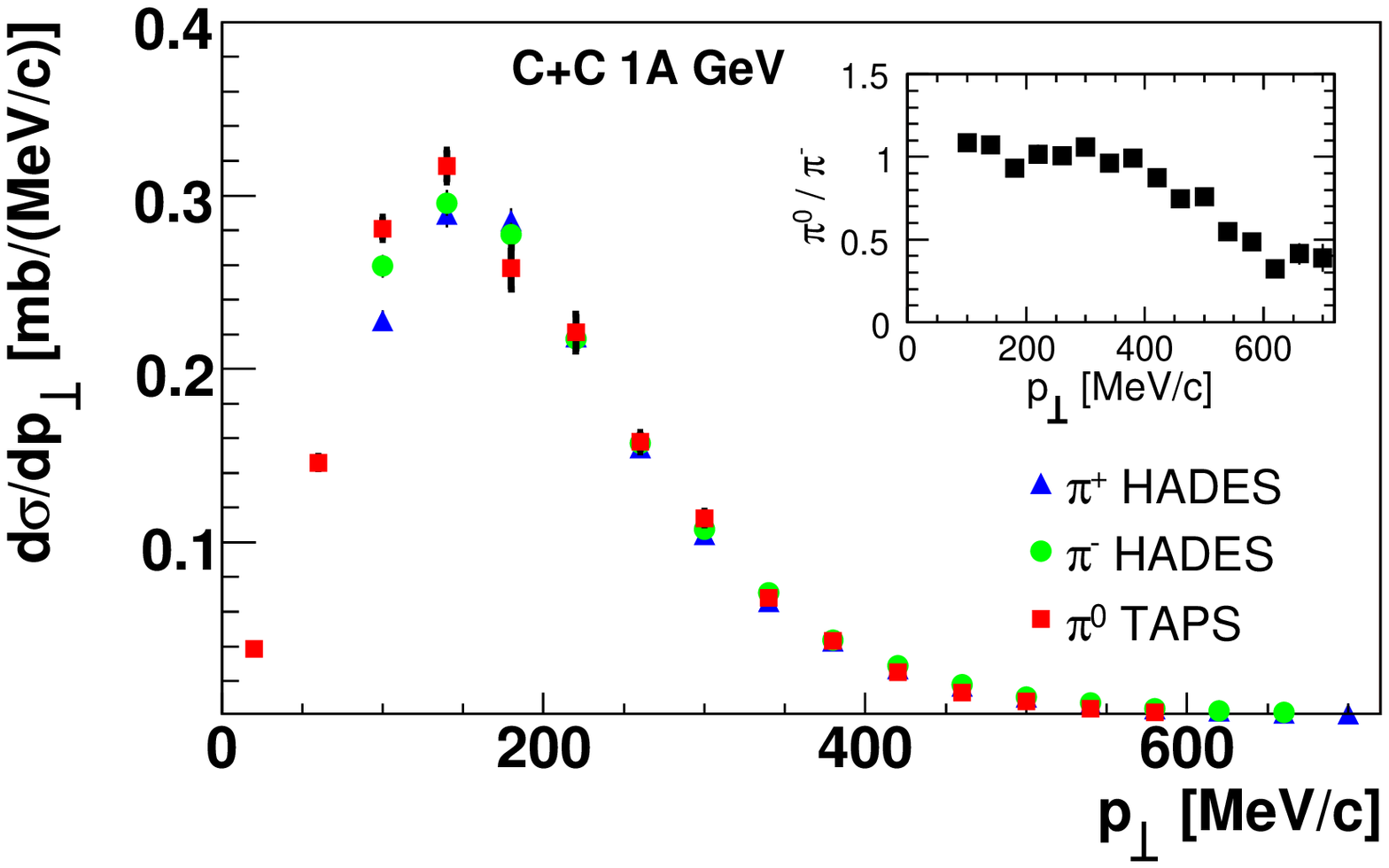}
  \includegraphics[width=1.\columnwidth]{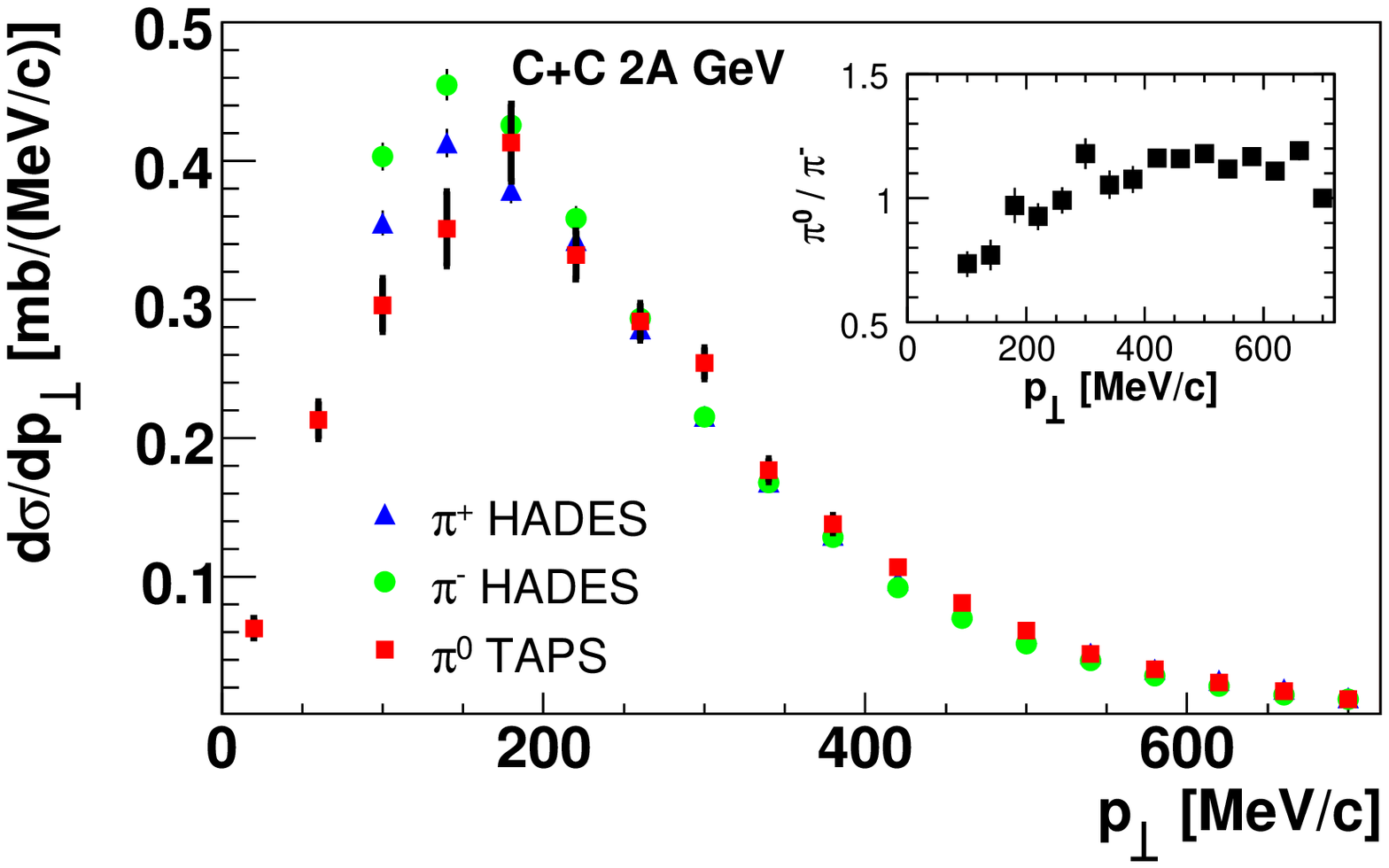}
  \includegraphics[width=1.\columnwidth]{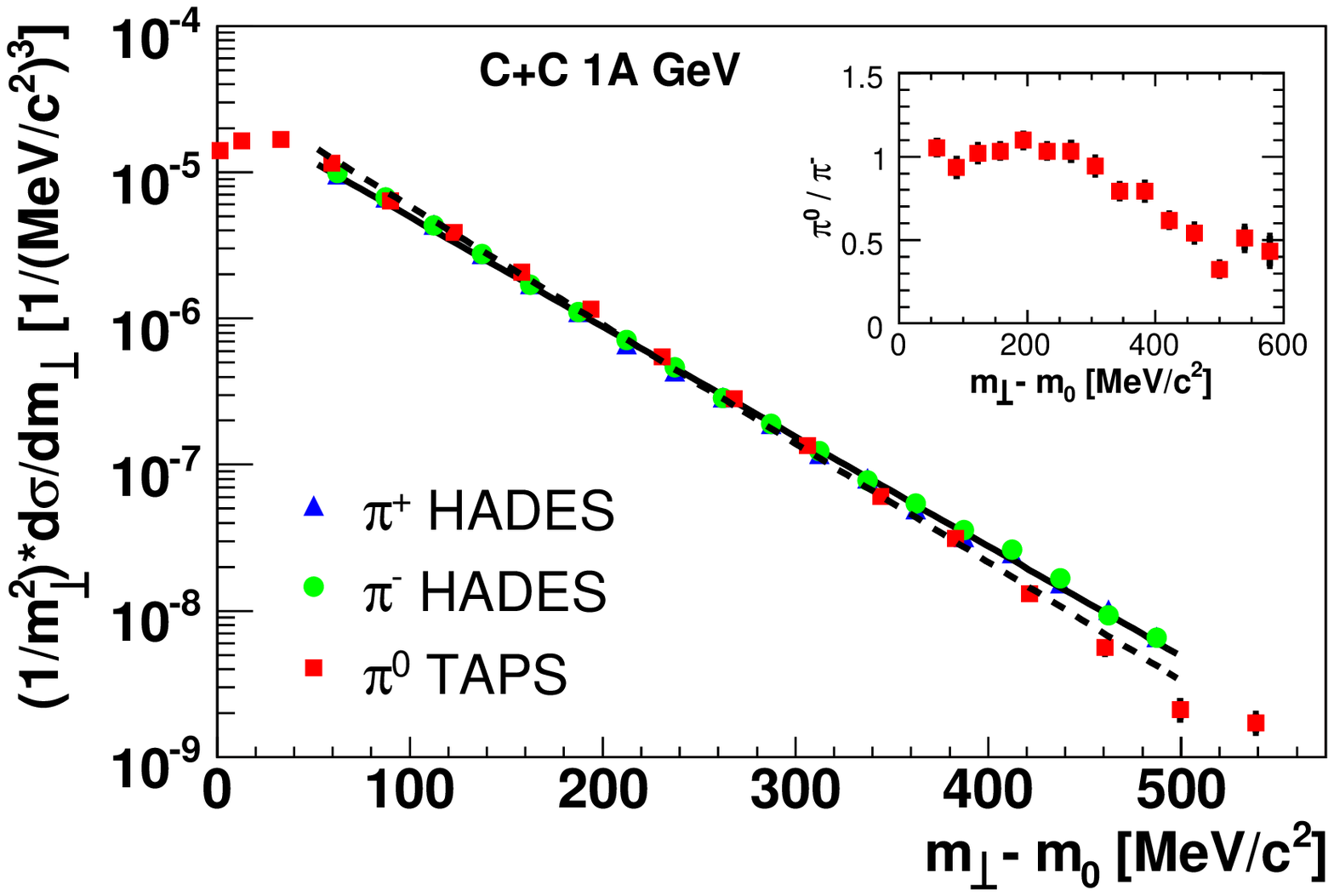}
  \includegraphics[width=1.\columnwidth]{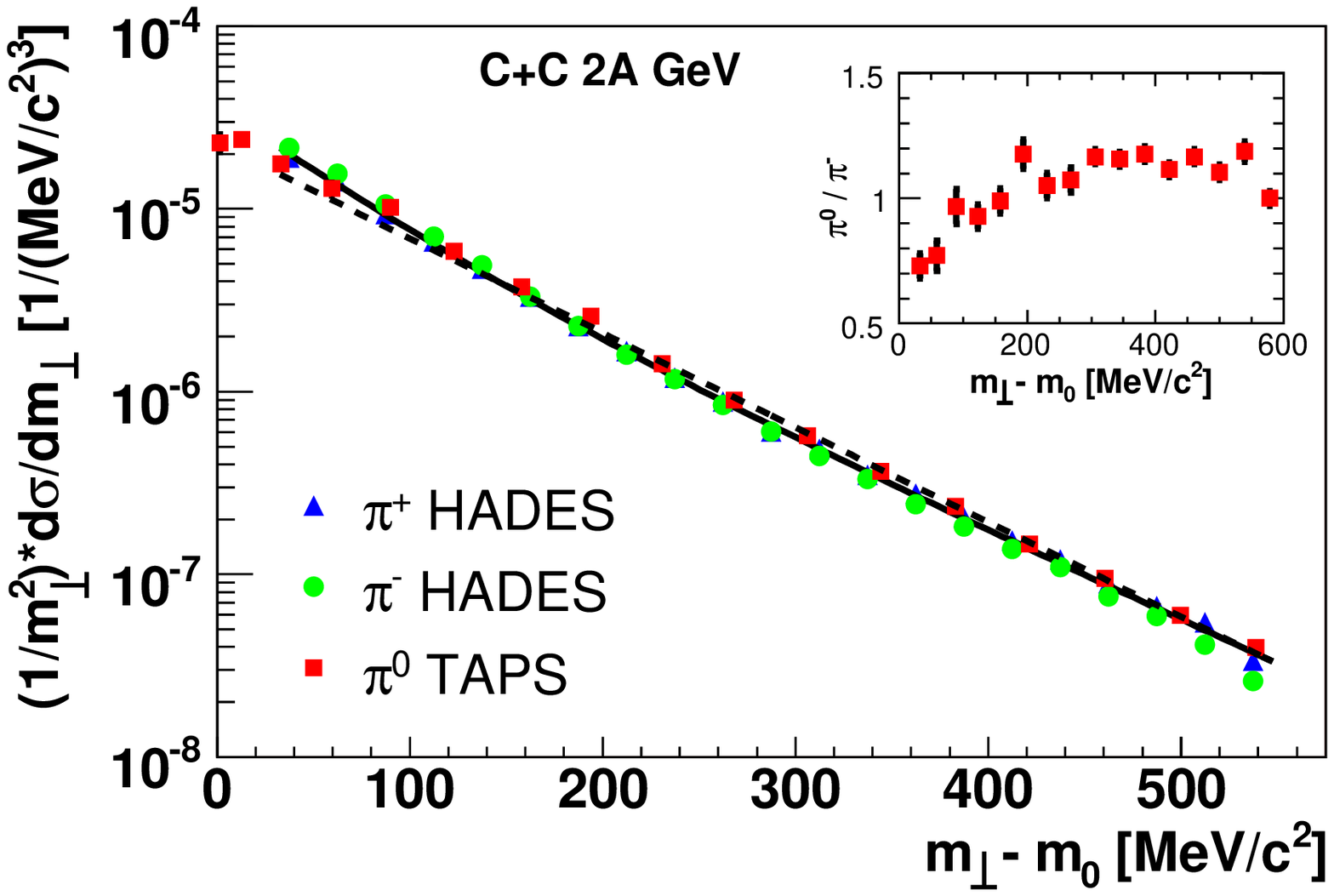}
  \includegraphics[width=1.\columnwidth]{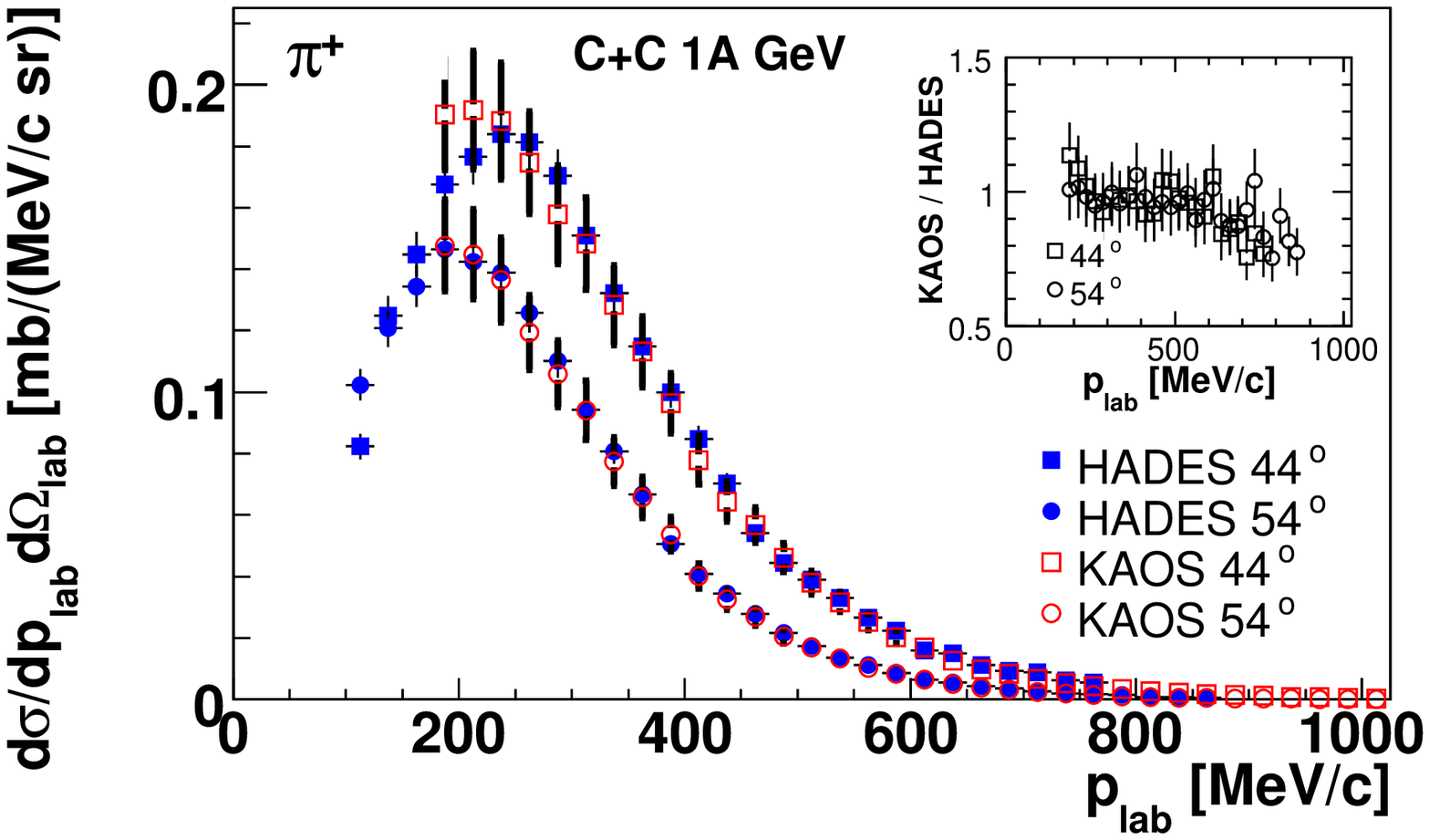}
  \hspace*{+.35cm}  
  \includegraphics[width=1.\columnwidth]{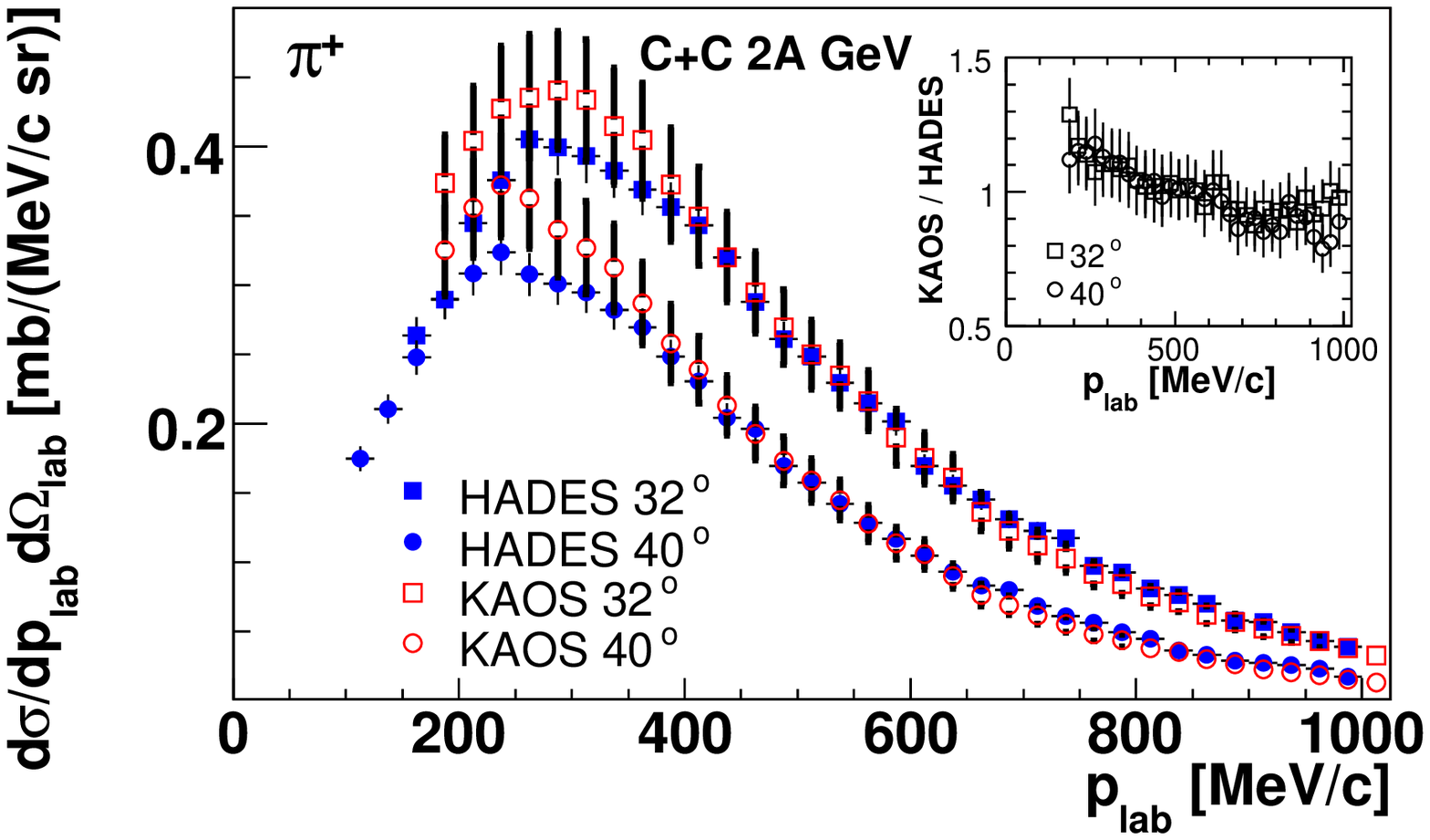}
  \caption[]{Comparison of transverse-mass and transverse-momentum 
  distributions 
  of $\pi$ mesons measured in the present experiment (HADES) and the previous 
  TAPS \cite{taps} and KaoS \cite{kaos} 
  experiments. Top: Transverse-momentum ($p_{\perp}$) 
  distributions of $\pi^{\pm,0}$ mesons for 1A GeV (left) and 2A GeV (right).
  Center: Transverse-mass ($m_{\perp}$) distributions of 
  $\pi^{\pm,0}$ mesons for 1A GeV (left) and 2A GeV (right). Full lines 
  show the results of fits of our data using two exponential 
  functions, while dashed lines show fits of the TAPS $\pi^{0}$ distributions 
  using one exponential function.
  Bottom: Momentum ($p_{lab}$) 
  distributions of $\pi^{+}$ mesons for 1A GeV (left) and 2A GeV (right).
  The HADES data were rescaled to minimum-bias cross sections and 
  filtered with the KaoS
  and TAPS acceptance filters (see text).
  Insets in all plots show the ratio of  $\pi$ yields on a linear scale.
 }
  \label{figtapshades}
\end{figure*}
\end{center}

\begin{table*}[htb]
\caption{Inverse slope parameters in units of MeV for $\pi^\pm$ measured at midrapidity 
derived from the
data (using one and two exponential functions) and UrQMD (using one exponential function) in
$^{12}$C~+~$^{12}$C collisions at 1~and~2A~GeV.}
\label{tab3}
\center~
\begin{tabular}{l c c c c c c}
\\
\hline
\multicolumn{7}{c}{Beam energy = 1A~GeV} \\
\hline
Particle & \multicolumn{2}{c}{Data} & \multicolumn{2}{c}{UrQMD} \\
 & T (1 slope)& $\chi^{2}/ndf$& T (1 slope)& $\chi^{2}/ndf$ \\
\hline
$\pi^+$ &57.8 $\pm$ 0.3 & 1.7 & 55.4 $\pm$ 0.3 & 2.2\\
$\pi^-$ & 57.9 $\pm$ 0.3 & 1.4 & 55.4 $\pm$ 0.3 & 2.0\\
\hline
\multicolumn{7}{c}{Beam energy = 2A~GeV} \\
\hline
Particle & \multicolumn{4}{c}{Data} & \multicolumn{2}{c}{UrQMD} \\
 & T (2 slopes)& $\chi^{2}/ndf$ & T (1 slope)& $\chi^{2}/ndf$& T (1 slope)& $\chi^{2}/ndf$ \\
\hline
$\pi^+$ & 47.7 $\pm$ 6.2; & 0.9 & 80.9 $\pm$ 0.5 & 4.7 & 86.5 $\pm$ 0.6 & 1.5\\
& 90.6 $\pm$ 3.3 \\

$\pi^-$ & 46.4 $\pm$ 5.2; & 1.2 & 76.7 $\pm$ 0.5 & 4.9 & 86.7 $\pm$ 0.6 & 1.4\\
& 84.4 $\pm$ 2.1 \\

\hline
\end{tabular}\\[3pt]
\end{table*}

Figures~\ref{dNdMpion_aug04rk} and \ref{dNdMpion} exhibit the measured and simulated
transverse mass distributions of $\pi^{+}$ and $\pi^{-}$ in different intervals of normalized rapidity $y_0 = (y_{lab}-y_{cms})/y_{cms}$
for $^{12}$C~+~$^{12}$C at 1A~GeV and 2A~GeV, respectively.
The systematic errors of the data are estimated from point-to-point  
differences
between distributions from the six independent HADES sectors as $\approx$ 5\%
(see Sect. 3.5). 

The transverse-mass ($m_{\perp}$) distributions have been fitted for each
rapidity bin using one
or two exponential functions.
The fit with two slopes employs
\begin{equation}
  \frac{1}{m_\perp^2} \frac{dN(y)}{dm_\perp} = 
  C_1(y) \, \exp\left(- \frac{m_\perp}{T_1(y)}\right) + 
  C_2(y) \, \exp\left(- \frac{m_\perp}{T_2(y)}\right)
\label{eq_mt}
\end{equation}
with $m_{\perp}=(p_{\perp}^2 + m^2)^{1/2}$, 
and $p_\perp$ as transverse momentum.
$C_{1,2}$ are normalizations and $T_{1,2}$ 
the inverse slope parameters; these parameters depend of course on rapidity.
For the 2A~GeV data, the two-component fit describes the experimental data 
better than a fit with one slope only (i.e., $C_2 \equiv 0$). 
For example, for the data at midrapidity $\chi^2/ndf$ is $\approx$ 1.0 
as compared to 4.8 for the fit with one slope. 

Figure~\ref{dNdMpion_aug04rk} clearly
demonstrates that for the lower bombarding energy of 1A~GeV, a fit
with one slope is sufficient for the description of the spectral
shape. The inverse-slope parameters for $\pi$ mesons  at
midrapidity for 1A~GeV $(-0.15 \le y_{0} \le 0.15)$ and 2A~GeV
$(-0.1 \le y_{0} \le 0.1)$ are summarized in Table~\ref{tab3} using
one or two exponential functions. The slopes of $\pi^+$ and
$\pi^-$ agree within error bars for the single exponential fit. 
At 2A~GeV, UrQMD predicts
different spectral shapes (purely exponential as compared to the concave shape 
of the real data), while at 1A~GeV
the agreement of UrQMD with data is better.
Note that the errors of the fitted parameters are much larger 
in the case of two-slope fits than for the single exponential fits. 
The reason is occurrence of large correlations between the fit parameters in the two-slope 
case, which cause a considerable increase of the errors. 

Thanks to the wide acceptance of the present experiment we can directly 
compare the data with results of
previous experiments on charged pion production in
$^{12}$C~+~$^{12}$C collisions at 1A~GeV and 2A~GeV done by the KaoS collaboration
\cite{kaos}, and for neutral pions at 1A~GeV and 2A~GeV obtained by the TAPS
collaboration \cite{taps}. For this purpose our data were passed through 
the respective acceptance filter
of the previous measurements: $\pm 4 ^{o}$ around the laboratory polar angles 
$\theta_{lab}$ of the KaoS setup, and $0.42<y_{lab}<0.74$ and 
$0.80<y_{lab}<1.08$ for 1A~GeV and 2A GeV, respectively, for the TAPS case. 
Then we rescaled the multiplicity per one LVL1 event to the 
minimum-bias cross section. 
For this we used a total reaction cross section of 0.95~b (calculated
according
to $\pi r_{0}^{2} (A^{1/3}_p +A^{1/3}_t)^{2}$,
assuming $r_{0} = 1.20~{\rm fm}$) and 
the ratio of $\langle $A$_{part} \rangle$
for minimum-bias and  LVL1 events from Table~\ref{tab4}.   
Fig.~\ref{figtapshades} shows a comparison of our pion transverse-mass 
and momentum 
distributions with those of the TAPS 
and KaoS experiments, respectively. 
The TAPS data are shown both in linear and logarithmic
scale to facilitate the comparison of yields as well as of spectral shapes.

It is apparent that the 
yields and spectral shapes measured by all three experiments are in general
fairly similar. The differences between integrated yields are within errors
and do not exceed 10\%.
Looking at Fig.~\ref{figtapshades} in more detail, one sees that the 
measured charged-pion data agree for both bombarding energies (lower two 
plots, see insets for data ratios) over a large momentum range
within errors (dominated by large systematic errors of the KaoS data). 
On the contrary, the comparison of
charged and neutral $\pi$ mesons (upper 4 plots) shows significant deviations
for both beam energies,
most apparent in the 2A GeV case (see inset of 
the upper right plot).
As a consequence, our data on charged pions can be 
described only by a two-slope fit,  
while the TAPS $\pi^{0}$ data at this energy need only  
a single-exponential fit \cite{taps}.
This may indicate differences in the reaction dynamics 
of charged and neutral $\pi$ mesons, not present in transport 
codes. It is interesting to note that the pattern of the 
$\pi^{0}/\pi^{-}$ ratio
changes with beam energy as well (see insets in middle row in 
Fig. \ref{figtapshades}).

Other fits of our pion transverse-momentum distributions are 
in principle conceivable.
For instance, a blast wave fit (see e.g. \cite{blast_fit})
can be used. However, if applied to only one particle type it does not 
allow for an unambiguous determination
of the flow parameter.  
  
\subsection{Rapidity distributions}

\begin{center}
\begin{figure*}[htb]
{
  \vspace*{+.1cm}
  \includegraphics[width=1.8\columnwidth]{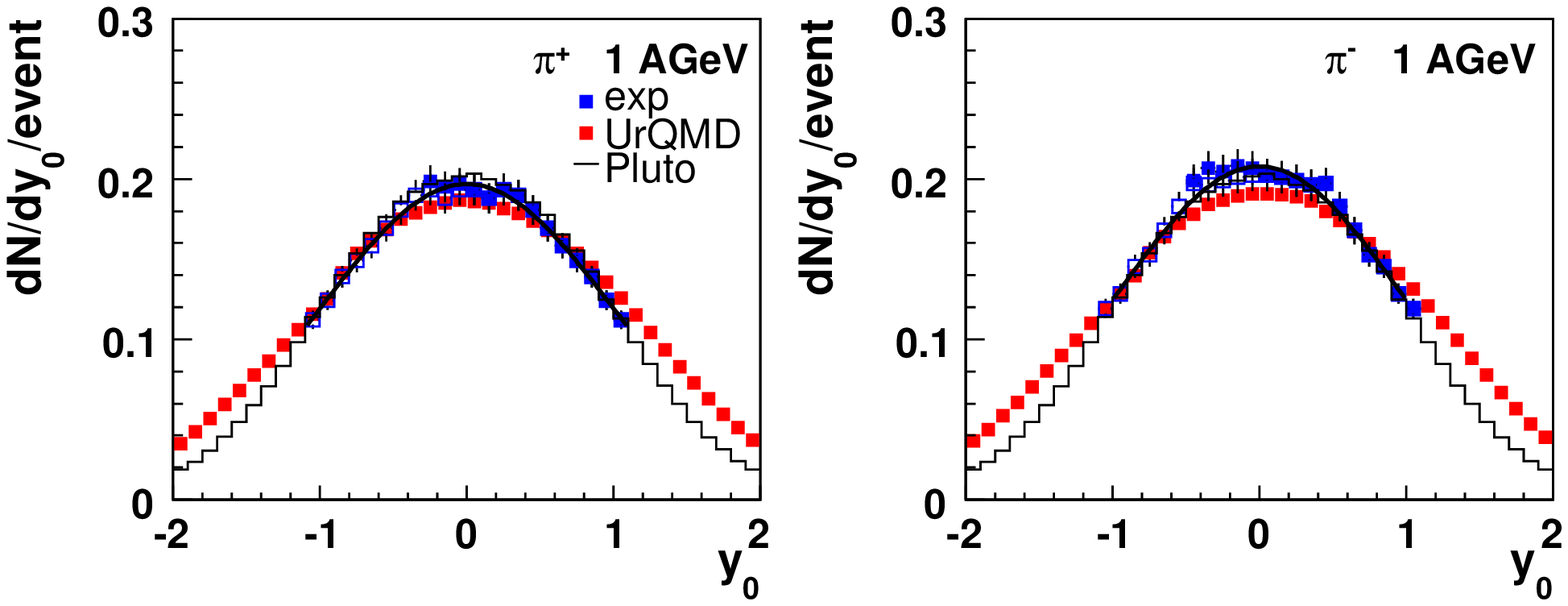}
  \vspace*{-0.1cm}
}
{
  \vspace*{+.5cm}
  \includegraphics[width=1.8\columnwidth]{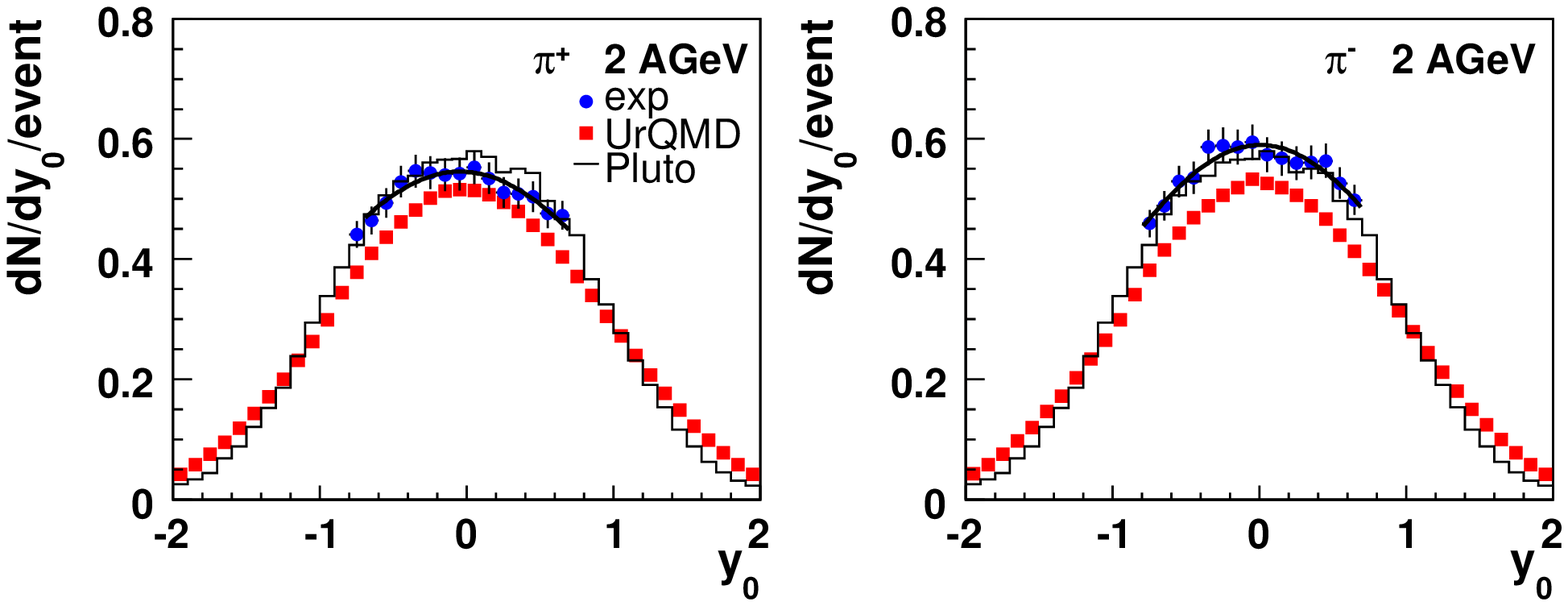}
  \vspace*{-0.1cm}
}
 \caption[]{The rapidity distribution of positively (left)
and negatively (right) charged pions produced in $^{12}$C~+~$^{12}$C collisions
at 1A~GeV (top) and 2A~GeV (bottom) for the LVL1 (semicentral) events. 
Circles with error bars show data, while full
squares depict UrQMD calculations. The distributions obtained with the
PLUTO generator are shown as histograms normalized to the measured yields 
(with $\pi^{+}$ and $\pi^{-}$ averaged). In the
1A~GeV case, the data points reflected from forward rapidities are
shown as open symbols. The slight asymmetries of $dN/dy$ with respect
to reflection at $y_0 = 0$ are used to check the systematic errors.}
\label{rapiditylab}
\end{figure*}
\end{center}

As seen in the previous section, the HADES acceptance in
transverse momentum
is rather large. For the missing parts of the acceptance at low and
high $p_{\perp}$ we extrapolated the pion yield. In doing so, for each
slice centered at rapidity $y$ the corresponding $p_{\perp}$
distribution was fitted with the function 
\begin{equation}
\frac{1}{p^2_\perp}\frac{dN}{dp_\perp} = 
c_1(y)  \exp\left(-\frac{p_\perp}{T_1(y)}\right) +
c_2(y)  \exp\left(-\frac{p_\perp}{T_2(y)}\right).
\end{equation}
The fit results were then used to integrate the $\pi^\pm$ yield outside
the acceptance. The resulting correction was $\approx 5\%$ and
$\approx 2\%$ for $\pi^+$ and $\pi^-$ respectively, except for
data at large rapidities, where the correction
was 10~-~20\%. Fig.~\ref{rapiditylab} shows the rapidity
distributions normalized to the number of LVL1 events obtained from
the  integration 
of the extrapolated
$p_{\perp}$ spectra of $\pi^{+}$ and $\pi^{-}$ from $^{12}$C~+~$^{12}$C 
at 1A~GeV and 2A~GeV. UrQMD simulation data (again normalized
to the number of LVL1 events) are
displayed as well. 
In order to extrapolate
our data outside the accepted rapidity range, we used the
Monte-Carlo event generator PLUTO
described above. The simulation parameters - inverse slopes and
anisotropies - were derived from our data, and 
the resulting rapidity distributions  
were normalized to the measured data.

The rapidity distributions exhibit a Gaussian-like shape with a standard
deviation ($\sigma$ referring to the scaled rapidity $y_0$) 
of about 1.0 for both beam energies. In the case of
the 1A~GeV data, the experimental rapidity distribution is about
20\% narrower than the one from UrQMD. This is in agreement with the
finding (see section 4.4 below) that the anisotropy parameter has a lower
value in the experimental data than in UrQMD simulation. Note that for
the 2A~GeV case, a slight underestimation of our data by UrQMD is observed.

\subsection{Multiplicities}

\begin{table}[htb]
\caption{Particle yields per reaction (LVL1 trigger condition) of
$\pi^\pm$ from $^{12}$C~+~$^{12}$C collisions. 
N$_\pi$ and N$_\pi$(4$\pi$) are the
measured and 4$\pi$ extrapolated yields, respectively.
The statistical
errors are negligible. Shown are the systematic errors due to the
various efficiency corrections
and to the 4$\pi$ extrapolation (see text for details).  \label{tab1}}
\center~
\begin{tabular}{c c c c}
\\
\hline
beam energy & particle &  $N_{\pi}$ & $N_{\pi}(4\pi$) \\
(A GeV) &  &  &  \\
\hline
1 & $\pi^+$ & 0.36$\pm$0.02 & 0.46$\pm$0.03$\pm$0.05 \\
1 & $\pi^-$ & 0.38$\pm$0.02 & 0.49$\pm$0.03$\pm$0.05 \\
\hline
2 & $\pi^+$ & 0.77$\pm$0.04  & 1.19$\pm$0.06$\pm$0.11 \\
2 & $\pi^-$ & 0.82$\pm$0.04  & 1.28$\pm$0.06$\pm$0.12 \\
\hline
\end{tabular}\\[3pt]
\end{table}

Pion yields ($N_{\pi}$) per reaction (under the LVL1 trigger condition) within
the HADES acceptance region as shown in Fig.~\ref{rapiditylab}
and in full phase space are presented in Table~\ref{tab1}.
The systematic error of the measured yield due to uncertainties 
in the detection/reconstruction/identification  efficiency is estimated 
as $5\%$, based again on a comparison of measurements in the six independent 
HADES sectors.
The extrapolation of the yields to full phase space is based on the
integration of the rapidity distributions simulated with PLUTO
(see above) and normalized to the data in the rapidity range covered
by HADES 
(see histograms in Fig.~\ref{rapiditylab}). 
Varying the input parameters of the PLUTO simulations - inverse slopes and 
angular anisotropy parameters - within their 
experimental errors, the differences between the rapidity distributions 
give us an estimate of the systematic error of the yield extrapolations of 
$9\%$.

Using the estimated averaged number of participants in the LVL1-triggered 
events (see in Table~\ref{tab4}), the resulting $\pi^\pm$ multiplicity per 
participant (averaged for $\pi^+$ and $\pi^-$) is then 
$0.055 \pm 0.003 \pm 0.005 \pm 0.004$ 
and $0.147 \pm 0.007 \pm 0.013~^{+0}_{-0.021}$ 
at 1A~GeV and 2A~GeV, respectively. 
The three systematic errors of the yields correspond to uncertainties connected
with the efficiency/purity corrections (5\%), the extrapolation to full
solid angle and full kinematic phase space (9\%), and the determination of
the number of participating nucleons (see Section 3.5).

Table~\ref{tab1b}
shows a comparison of measured $\pi$ meson
multiplicities per participant and results of our UrQMD simulations as well.
The results agree within quoted errors.

\begin{table}[htb]
\caption{Comparison of multiplicities per participant of $\pi$ mesons 
derived from our data with UrQMD results.\label{tab1b}}
\center~
\begin{tabular}{c c c c}
\\
\hline
beam energy & particle & this work & UrQMD \\
(A GeV) &  & $\times 10^{-3}$ & $\times 10^{-3}$ \\
\hline
1 & 1/2($\pi^+$+$\pi^-$) & $55 \pm 3 \pm 5 \pm 4$  & 59\\
1 & $\pi^0$ &  & 67\\
\hline
2 & 1/2($\pi^+$+$\pi^-$) & $147 \pm 7 \pm 13~^{+0}_{-21}$  & 137\\
2 & $\pi^0$ & & 159\\
\hline
\end{tabular}\\[3pt]
\end{table}

As we have shown above by comparing the momentum distributions, our 
results on integrated charged pion yields are consistent with the 
TAPS $\pi^{0}$ as well as the KaoS $\pi^{+}$ data. 
Due to experimental errors, it is however difficult to draw
a conclusion on the difference in production yields of neutral and charged
$\pi$ mesons predicted by UrQMD (see Table \ref{tab1b}).

\subsection{Angular distributions}

\begin{center}
 \begin{figure*}[htb]
{
  \vspace*{+.1cm}
  \includegraphics[width=1.8\columnwidth]{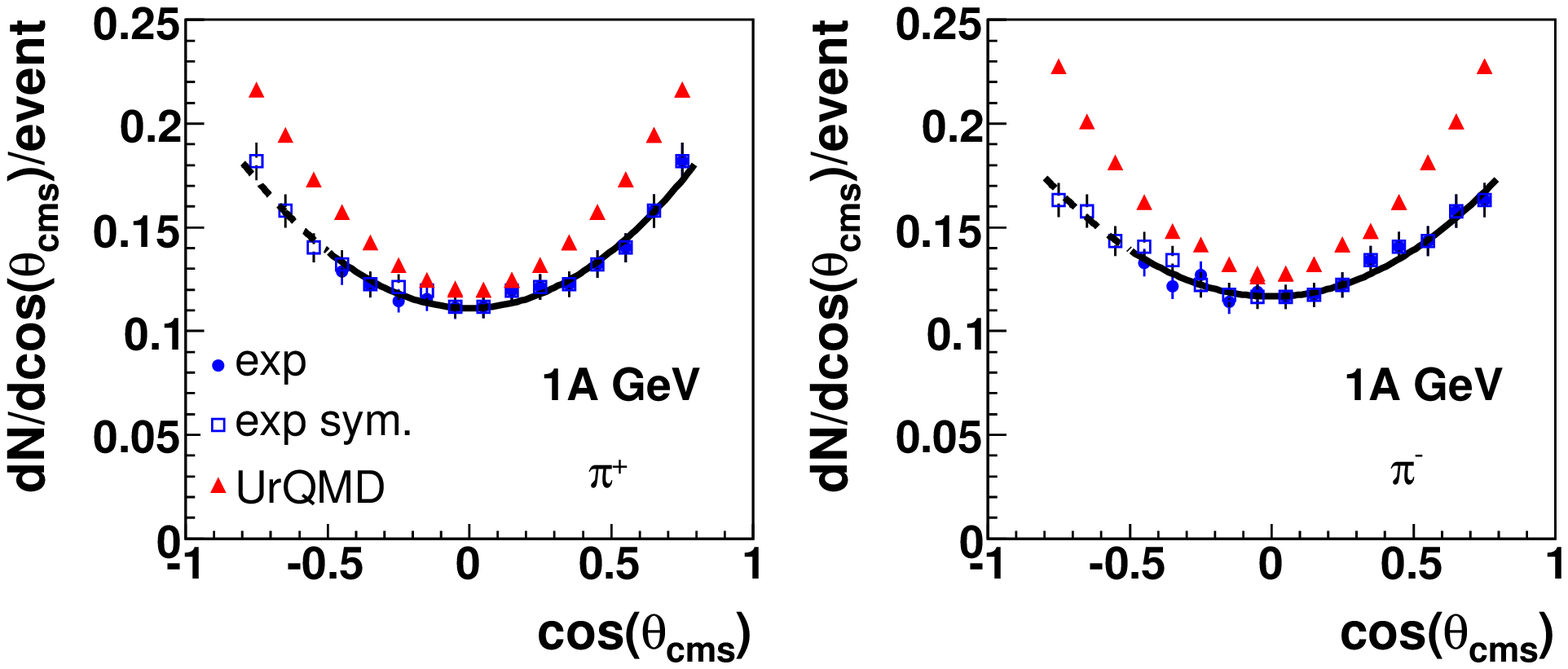}
  \vspace*{-0.1cm}
}
{
  \vspace*{+.5cm}
  \includegraphics[width=1.8\columnwidth]{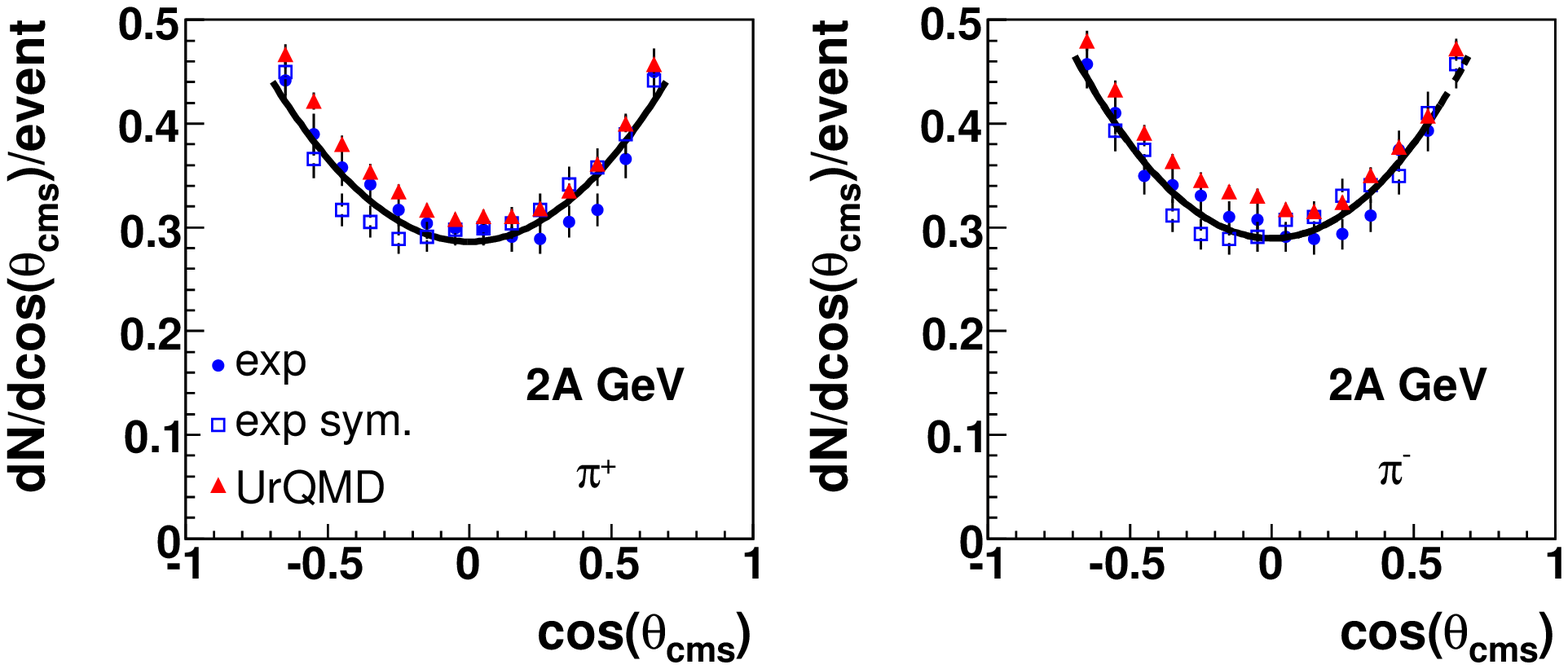}
  \vspace*{-0.1cm}
}

 \caption[]{Polar angle distribution in the center-of-mass system of
   positively (left) and negatively (right) charged $\pi$ mesons produced in
   $^{12}$C~+~$^{12}$C collisions at 1A~GeV (top) and 2A~GeV (bottom)
   for the LVL1 (semicentral) events.
   Pions with center-of-mass momenta 200~-~800~MeV/c have been selected.
   Full circles show measured data. The data points reflected around $90^\circ$
   are shown as open squares. The full lines show the fit by 
   Eq. (\ref{eq:A2eq}), the extrapolations of the fits outside the 
   acceptance are plotted as dashed lines. The UrQMD points are displayed as 
   triangles, no fits to the UrQMD  points are shown.
}
  \label{pid4cm}
\end{figure*}
\end{center}

The measured centre-of-mass polar angular distributions of pions
produced in 1A~GeV and 2A~GeV $^{12}$C~+~$^{12}$C collisions are exhibited in Fig.~\ref{pid4cm}, together with the corresponding UrQMD distribution.
Pions with centre-of-mass momenta between 200 and 800 MeV/c have been selected.
No losses in acceptance occur in
this phase space region for the range of polar angles 
shown in Fig.~\ref{pid4cm}.
The systematic errors of the data are again estimated from the point-to-point 
differences between distributions from the six independent HADES sectors 
to be $\approx$ 5\% (see Sect. 3.5). 
As seen from the comparison of the measured data points and the ones
reflected around $90^\circ$ (see Fig. ~\ref{pid4cm}), this error
underestimates the observed differences between the forward and
backward hemisphere by a factor of $<$ 1.5.

In the symmetric collision system $^{12}$C~+~$^{12}$C the polar 
distributions in the
center-of-mass system can be fitted with the following expression
\begin{equation}
\frac{dN}{d (\cos \theta_{cms}) } =  A_1 (1 + A_2 \cos^2 \theta_{cms} ).
\label{eq:A2eq}
\end{equation}
The fit parameter $A_2$ characterizes the anisotropy of the pion
source, and $A_1$ is a normalization. As visible in Fig.~\ref{pid4cm},
the data show strong anisotropies quantified by 
$A_2 = 0.88 \pm 0.12$ and $1.19 \pm 0.16$
for beam energies of 1A~GeV and 2A~GeV, respectively. The UrQMD model
gives similar anisotropies with $A_2$ = 1.45 and $A_2$ =
1.12 ($A_2$ = 0.56 and $A_2$ = 0.70 when integrated over all momenta,
including also the region outside our acceptance). 

\begin{center}
 \begin{figure*}[htb]
{
  \vspace*{+.01cm}
  \includegraphics[width=1.8\columnwidth]{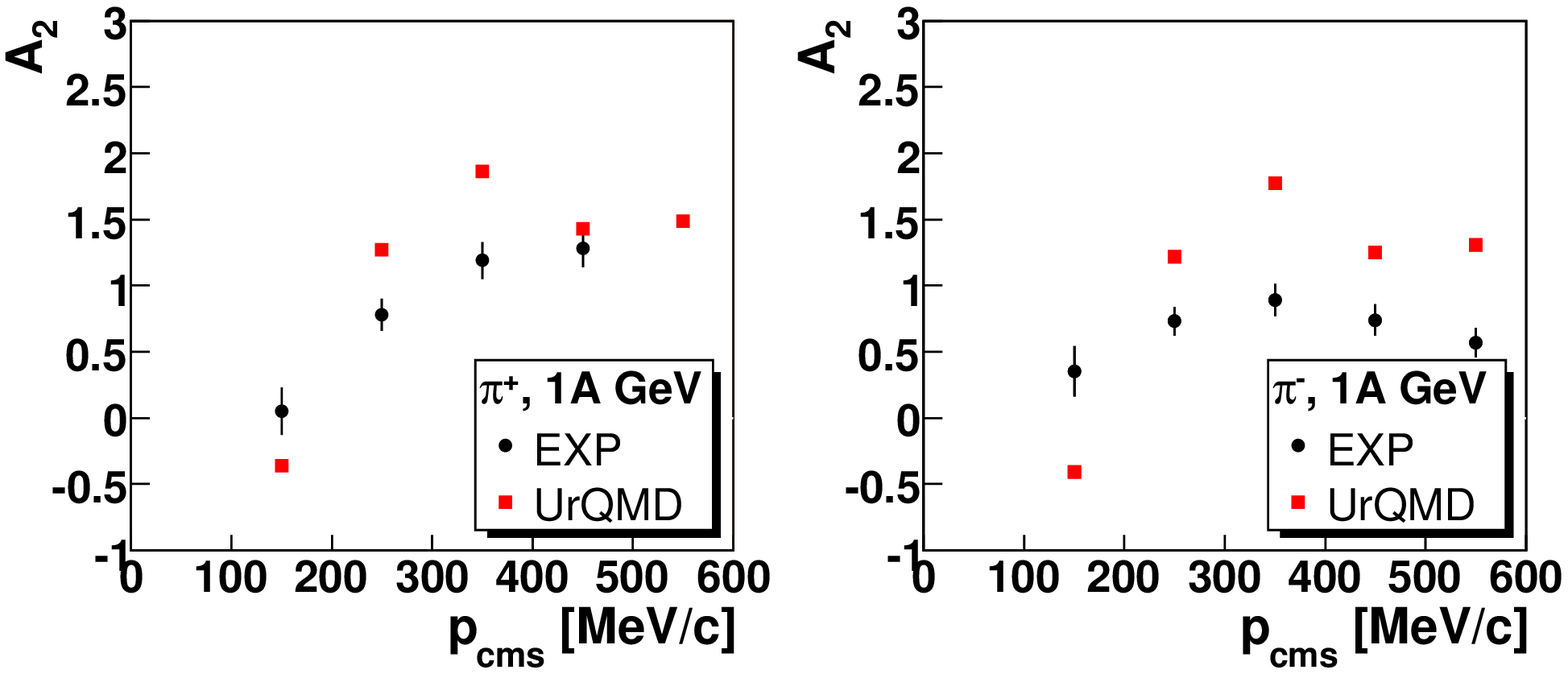}
  \vspace*{-0.1cm}
}
{
  \vspace*{+.5cm}
  \includegraphics[width=1.8\columnwidth]{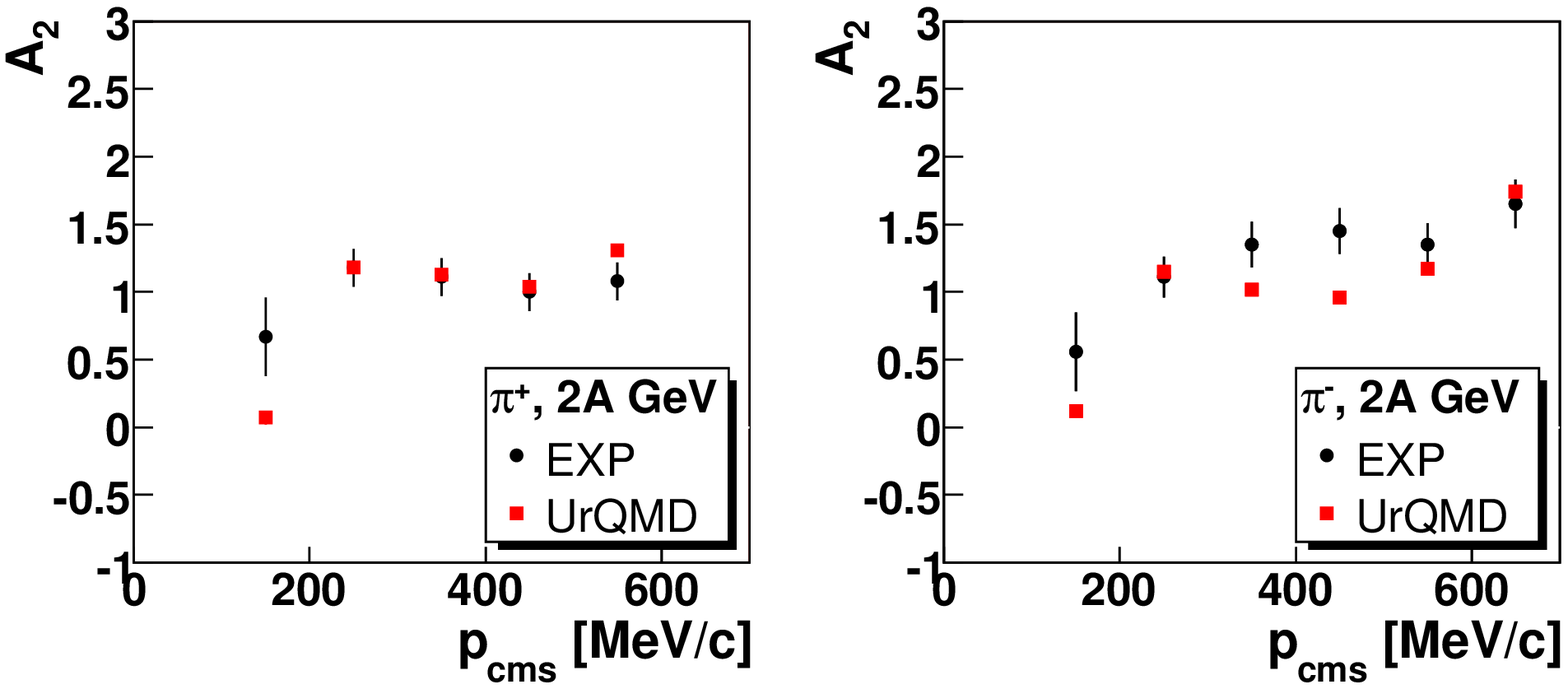}
  \vspace*{-0.1cm}
}
\caption[]{Dependence of the anisotropy parameter $A_2$ on the momentum in
  the center-of-mass system for positively (left) and negatively (right) 
  charged pions
  produced in $^{12}$C~+~$^{12}$C collisions at 1A~GeV (top) and 2A~GeV 
  (bottom) for the LVL1 (semicentral) events.
  Circles with error bars show fits to the data, while squares exhibit fits to 
  UrQMD simulations after they were subjected to the full analysis chain. 
  Errors of 
  the UrQMD points are smaller than the symbol size.
}
  \label{pid4cm_mom}
\end{figure*}
\end{center}

We observe a strong dependence of the
anisotropy on momentum. This is evident from Fig.~\ref{pid4cm_mom},
where $A_2$ is displayed as a function of the pion's centre-of-mass momentum 
from fits of Eq. (\ref{eq:A2eq}) to our data.
The data points for $\pi^{+}$ at highest displayed momenta are not shown, 
as they correspond to the phase space region ($p_{lab}$ approaching 1000 MeV/c)
where reliable 
particle identification becomes gradually impossible (see Fig. \ref{pid2} in
Sect. 3.3).

One can see that the anisotropy steadily increases with momentum for
both pion charges and both beam energies up to $A_2 \simeq 1.0 -
1.5$, around~400 MeV/c, where it has a tendency to level off. This
behavior is fairly well reproduced by the UrQMD model, which,
however, tends to level off at somewhat larger values of the anisotropy
in case of the 1A GeV data (see Fig.~\ref{pid4cm_mom}).
For the data in the region 100~MeV/c $< p_{cms} <$ 200~MeV/c, the 
anisotropy can be fitted in a limited $\Theta_{cms}$ range only. 
The results show significantly lower $A_2$ parameters for both 
the data and UrQMD. The close-to-zero (at 2A~GeV) and even slightly 
negative anisotropies given by UrQMD for low momenta at 1A~GeV are 
not seen in the experiment.

It is interesting to compare these distributions to the
corresponding ones of the elementary $NN \rightarrow N\Delta \to NN \pi$ 
reactions
which are expected to be the dominant source of pion production at
these beam energies \cite{Mosel1,Mosel2}. In order to estimate the latter
we have again used our PLUTO
generator employing measured $\Delta$ distributions to model the 
proton-proton reaction \cite{ppdelta}. 
We find that the shape of the simulated $A_2$ distributions is similar
to the one found in $^{12}$C~+~$^{12}$C, but
$A_2$ levels off at substantially higher
values: namely 3.5 and 5 for 1A~GeV and~2A~GeV, respectively.
The lower $A_2$ values found for $^{12}$C~+~$^{12}$C may be attributed to pion
re-scattering and final state interactions even in the small
system under consideration here.  

Prior to the present work no data had been published on pion anisotropies
in $^{12}$C~+~$^{12}$C collisions.  Early studies of pion production in
Ne-induced reactions at 0.8A~GeV on NaF, Cu and Pb targets had found almost
isotropic angular distributions for very low pion cms kinetic energies
($E_{\pi^+} \le 50$~MeV), but substantial anisotropies for all higher
energies ($E_{\pi^+} \ge 150$~MeV) \cite{chiba}. The two closest systems
studied most comprehensively in this respect are Ar+KCl at 0.8A~GeV as well as
1.8A~GeV \cite{brockman,nagamiya} and Ca+Ca at 1.93A~GeV \cite{reisdorf}.
While in both cases similar momen\-tum-averaged anisotropies were observed, 
with values
of $\langle A_2 \rangle = 0.5 - 0.6$, only the Ar+KCl data display a strong 
pion-energy
dependence of $A_2$, peaking around $E_{\pi}$ = 200 - 300 MeV.  Based on a
comparison with transport calculations, the authors of Ref.~\cite{reisdorf}
attributed these differences to the very different centralities covered by
the two measurements.  As seen from Fig.~\ref{pid4cm_mom} and as discussed
above, we do not observe in ${}^{12}$C~+~${}^{12}$C, 
at both bombarding energies, a rise and
fall of $A_2$, but rather a simple levelling-off with increasing pion
cms momenta.

As the momentum dependence of $A_2$ has been claimed 
to depend strongly on the 
reaction centrality \cite{reisdorf}, we tried to study this effect in 
our data sample.
First we checked, how the distribution differs in UrQMD events for the 
mini\-mum-bias and LVL1-triggered events (see Sect. 3.4). The effect is 
negligible, as could be expected for a very weak preference of more central
events of the LVL1 trigger in the light $^{12}$C~+~$^{12}$C system.
Then we selected the 10\% of ``most central'' events in the measured
and UrQMD data samples by requiring more than 3 identified protons in the event.
Again, we did not observe any significant difference with respect to 
the LVL1 data samples.

\section{Summary}

In summary, the charged-pion characteristics in the reaction
${}^{12}$C~+~${}^{12}$C at 1A~GeV and 2A~GeV have been measured 
in detail with the HADES
spectrometer. The found $\pi$ meson yields are in good
agreement with previous results obtained with
the TAPS and KaoS detectors. The much larger acceptance of
the present experiment, as compared to older measurements allowed for 
a more precise 
and reliable extrapolation of the yields to full solid angle, which is 
essential for the normalization of our dielectron yields
in this reaction system. 

Our data on the pion transverse-mass distributions at midrapidity 
can be described by a Maxwell-Boltzmann function
for 1A~GeV, while the 2A~GeV data show a strong second exponential
component
with smaller slope. This finding is in agreement with former results
for the same system and energy for $\pi^\pm$ \cite{kaos}, whereas
$\pi^0$ data \cite{taps} exhibit only a one-slope behaviour. 
The reasonable agreement of our transverse-momentum
spectra at 1A~GeV with UrQMD calculations indicates that the degree 
of thermalization in the light ${}^{12}$C~+~${}^{12}$C system is 
adequately reproduced in this transport model. At 2A~GeV, however,
some deviations of our data from the UrQMD results become visible.

In contrast to the TAPS and KaoS 
data, which were measured in a limited angular range,
strong pion anisotro\-pies could be
observed in the much larger acceptance region of the present experiment.
The systematics \cite{reisdorf} of pion production in heavier systems
at comparable beam energies points to similar anisotropy values as extracted
from our data sample. The asymmetries have a non-trivial momentum dependence.
A peaking of the asymmetry as reported in \cite{brockman}, however,
can not be confirmed.
The comparison of our results with  proton-proton data 
suggests that the angular anisotropies observed in  ${}^{12}$C~+~${}^{12}$C
are remnants of the much stronger effect characteristic
for inelastic nucleon-nucleon scattering.

\section*{Acknowledgments} 

The HADES collaboration gratefully
acknowledges the support by BMBF grants 06MT238, 06TM970I, 06GI146I, 
06F-140, 06FY171, and 06DR135, 
by DFG EClust 153 (Germany), 
by GSI (TM-KRUE, TM-FR1, GI/ME3, and OF/STR), 
by grants GA AS CR IAA100480803 and MSMT LC 07050 (Czech Republic), 
by grant KBN 5P03B 140 20 (Poland), 
by INFN (Italy), by CNRS/IN2P3 (France), by 
grants~~ MCYT~~ FPA2006-09154,~~ XUGA PGID IT06PXIC296091PM 
and CPAN CSD2007-00042 (Spain), 
by grant FTC POCI/FP /81982 /2007  (Portugal),
by grant UCY-10.3.11.12 (Cyp\-rus), by INTAS grant
06-1000012-8861 and EU contract RII3-CT-2004-506078.


\end{document}